\documentclass[aps,prd,onecolumn,nofootinbib,superscriptaddress,preprintnumbers,balancelastpage,longbibliography,nobibnotes]{revtex4-1}
\usepackage{graphicx,epsfig,psfrag,bm,amssymb}
\usepackage{dcolumn}
\usepackage{bm}
\usepackage[dvipsnames]{xcolor}
\usepackage{braket}
\usepackage{mathrsfs,amsfonts,color}
\usepackage{hepunits}
\usepackage{hyperref}
\usepackage{comment}
\usepackage{soul}
\usepackage[caption=false]{subfig}

\definecolor{linkcolor}{rgb}{0.0, 0.28, 0.67}
\hypersetup{colorlinks=true,
linkcolor=black,
citecolor=linkcolor,
urlcolor=linkcolor,
linktocpage=true
}

\DeclareRobustCommand{\Eq}[1]{Eq.~(\ref{#1})}

\newcommand{\Sec}[1]{Sec.~\ref{sec:#1}}
\newcommand{\App}[1]{Appendix~\ref{app:#1}}
\newcommand{\Fig}[1]{Fig.~\ref{fig:#1}}

\newcommand{\GeV}{\text{GeV}}
\newcommand{\GHz}{\text{GHz}}
\newcommand{\kHz}{\text{kHz}}
\newcommand{\Hz}{\text{Hz}}
\newcommand{\cm}{\text{cm}}

\newcommand{\be}{\begin{eqnarray}}
\newcommand{\ee}{\end{eqnarray}}
\newcommand{\E}{\mathbf{E}}
\newcommand{\B}{\mathbf{B}}

\newcommand{\Uv}{\mathbf{U}}

\newcommand{\n}{\mathbf{n}}
\newcommand{\x}{\mathbf{x}}

\newcommand{\K}{\mathbf{K}}

\newcommand{\jeff}{j_\text{eff}}

\newcommand{\order}[1]{\mathcal{O}{(#1)}}
\newcommand{\nl}{\nonumber \\}
\newcommand{\w}{\omega}
\newcommand{\wg}{\omega_g}
\newcommand{\wLC}{\omega_\text{LC}}

\newcommand{\U}{{\mathbf U}}
\newcommand{\A}{{\mathbf A}}
\newcommand{\grad}{\nabla}

\newcommand{\pd}{\partial}

\newcommand{\xv}{{\bf x}}
\newcommand{\Vcav}{V_\text{cav}}
\newcommand{\Vshell}{V_\text{shell}}
\newcommand{\VLC}{V_\text{LC}}
\newcommand{\Lcav}{L_\text{cav}}
\newcommand{\Mcav}{M_\text{cav}}

\newcommand{\p}{\varphi}
\newcommand{\Lop}{\mathbf{L}}

\newcommand{\hpd}{h^\text{PD}}
\newcommand{\htt}{h^\text{TT}}
\newcommand{\htthat}{\hat{h}^\text{TT}}

\newcommand{\jv}{\boldsymbol{j}}

\newcommand{\f}{{\bf f}}

\usepackage[page,toc,titletoc,title]{appendix}

\begin{document}

\preprint{FERMILAB-PUB-22-892-SQMS-T}

\title{
%Make MAGO Great Again: 
MAGO\,2.0\,: Electromagnetic Cavities as Mechanical Bars for Gravitational Waves
}

\author{Asher Berlin}
\affiliation{Theoretical Physics Division, Fermi National Accelerator Laboratory, Batavia, IL 60510, USA}
\affiliation{Superconducting Quantum Materials and Systems Center (SQMS), Fermi National Accelerator Laboratory, Batavia, IL 60510, USA}
\author{Diego Blas}
\affiliation{Grup de F\'{i}sica Te\`{o}rica, Departament de F\'{i}sica, Universitat Aut\`{o}noma de Barcelona, 08193 Bellaterra, Spain}
\affiliation{Institut de Fisica d’Altes Energies (IFAE), The Barcelona Institute of Science and Technology, Campus UAB, 08193 Bellaterra (Barcelona), Spain}
\author{Raffaele Tito D’Agnolo}
\affiliation{Universit\'e Paris-Saclay, CEA, Institut de Physique Th\'eorique, 91191, Gif-sur-Yvette, France}
\author{Sebastian~A.~R.~Ellis}
\affiliation{D\'epartement de Physique Th\'eorique, Universit\'e de Gen\`eve, 
24 quai Ernest Ansermet, 1211 Gen\`eve 4, Switzerland}
\author{Roni Harnik}
\affiliation{Theoretical Physics Division, Fermi National Accelerator Laboratory, Batavia, IL 60510, USA}
\affiliation{Superconducting Quantum Materials and Systems Center (SQMS), Fermi National Accelerator Laboratory, Batavia, IL 60510, USA}
\author{Yonatan Kahn}
\affiliation{Department of Physics, University of Illinois Urbana-Champaign, Urbana, IL 61801, USA}
\affiliation{Illinois Center for Advanced Studies of the Universe, University of Illinois Urbana-Champaign, Urbana, IL 61801, USA}
\affiliation{Superconducting Quantum Materials and Systems Center (SQMS), Fermi National Accelerator Laboratory, Batavia, IL 60510, USA}
\author{Jan Sch\"{u}tte-Engel}
\author{Michael Wentzel}
\affiliation{Department of Physics, University of Illinois Urbana-Champaign, Urbana, IL 61801, USA}
\affiliation{Superconducting Quantum Materials and Systems Center (SQMS), Fermi National Accelerator Laboratory, Batavia, IL 60510, USA}

\begin{abstract}
%Make MAGO great again!
Superconducting cavities can operate analogously to Weber bar detectors of gravitational waves, converting mechanical to electromagnetic energy. The significantly reduced electromagnetic noise results in increased sensitivity to high-frequency signals well outside the bandwidth of the lowest mechanical resonance. In this work, we revisit such signals of gravitational waves and demonstrate that a setup similar to the existing ``MAGO" prototype, operating in a scanning or broadband manner, could have sensitivity to strains of $\sim 10^{-22} - 10^{-18}$ for frequencies of $\sim 10 \ \text{kHz} - 1 \ \text{GHz}$.
\end{abstract}

\maketitle

\tableofcontents

\newpage
\section{Introduction} 

The first detection of gravitational waves (GWs) in the Hz to kHz range by the LIGO and Virgo collaborations~\cite{LIGOScientific:2016aoc} was the pinnacle of decades of research activity, opening a new path to observe the Universe. The science case to extend the observational frequency range is especially strong. This has motivated the current landscape for existing and planned efforts, including CMB~\cite{Namikawa:2019tax,CMB-S4:2020lpa} and pulsar timing array~\cite{NANOGrav:2020bcs,Janssen:2014dka} measurements, future laser~\cite{Hild:2010id, Punturo:2010zz,LIGOScientific:2016wof,LISA:2017pwj,Yagi:2011wg} and atom interferometers~\cite{Badurina:2019hst,Abe:2021ksx,AEDGE:2019nxb}, and new astrophysical signatures~\cite{Blas:2021mqw,Fedderke:2021kuy, Fedderke:2022kxq}, which together hold promise to  cover a wide range of frequencies below a kHz. There is also motivation to extend sensitivity to higher frequencies. While searches with small correlated interferometers~\cite{Holometer:2016qoh} and piezoelectric mechanical resonators~\cite{Goryachev:2021zzn} have produced initial sensitivity focused around a MHz, there are no established methods to systematically cover the orders of magnitude of the unexplored GW spectrum ranging from kHz to GHz, which is the focus of this work. 

Aside from a few exceptions~\cite{Ghiglieri:2015nfa,Ghiglieri:2020mhm,Ringwald:2020ist,Ghiglieri:2022rfp,Casalderrey-Solana:2022rrn}, high-frequency GWs are harbingers of physics beyond the Standard Model both in the cosmos today and the earliest stages of the Universe, as no known astrophysical objects are sufficiently dense to produce GWs above $\sim 10 \ \text{kHz}$. Perhaps the most well-motivated of such signals are those generated by primordial cosmological events. Indeed, causality restricts their wavelength to be smaller than the Hubble radius at the time of production, which implies that GWs originating from when the universe was hotter than $10^{11} \ \GeV$ are guaranteed to have a frequency above $10 \ \kHz$ today. Unfortunately, measuring such primordial GWs is extremely challenging, since the successful predictions of Big Bang nucleosynthesis and measurements of the CMB severely restrict their contribution to the radiation energy density (see, e.g., Ref.~\cite{Pagano:2015hma}). This puts the direct observation of primordial GWs above $\sim 1 \ \kHz$ out of reach for present-day experiments. Other new physics sources of high-frequency GWs may result in larger signals, but these are less universal, such as those arising from the inspiral of light primordial black holes (PBHs)~\cite{Franciolini:2022htd,Franciolini:2022ewd} or the superradiant production and annihilation of light bosons around nearby black holes~\cite{Brito:2015oca,Arvanitaki:2010sy,Arvanitaki:2014wva}; while slightly less generic, both would point to exciting new physics if discovered. 

In this work, we discuss an experimental strategy capable of detecting these more speculative signals. The approach we consider, based on the interaction between electromagnetic (EM) and mechanical resonances, was pioneered in the late 1970s~\cite{Braginskii:1973vm,Pegoraro:1978gv,Pegoraro_1978} and led to a nascent experimental effort by the ``MAGO" (Microwave Apparatus for Gravitational Waves Observation) collaboration in the early 2000s~\cite{Ballantini:2003nt,Ballantini:2005am} which was unfortunately culled before coming to fruition (a prototype version of this experiment is currently in public display at the University of Genoa). In our study, we combine an improved understanding of the signal and noise sources in a prototypical setup to argue for reviving interest in this experimental program. 

\begin{figure}[t]
\centering
\includegraphics[width=0.7\textwidth]{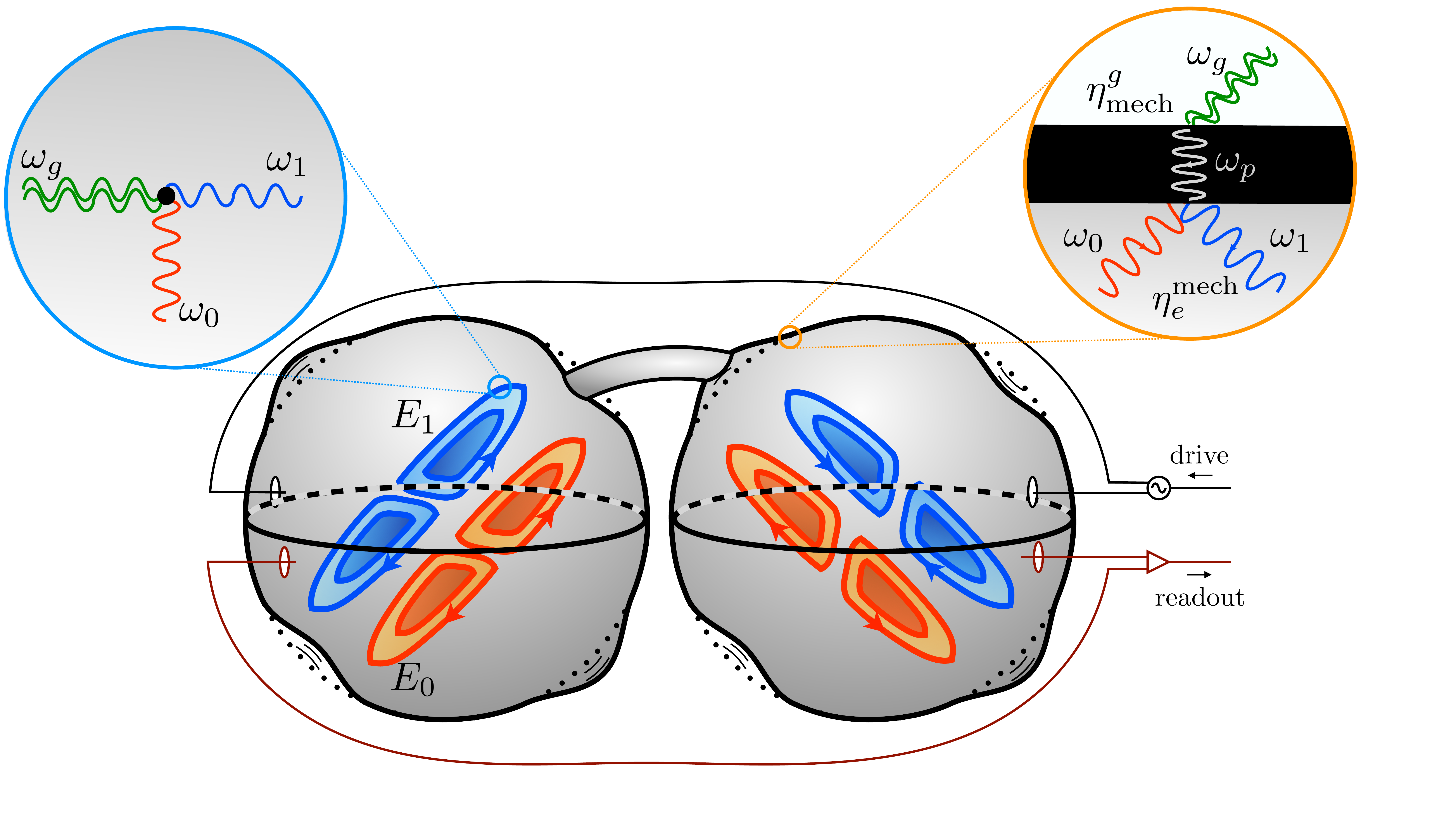}
\caption{Cartoon of a two-spherical-cell setup, illustrating the two coexisting signals. The pump mode $E_0$ of the cavity is driven at frequency $\w_0 \sim 1 \ \GHz$ (orange). The incoming gravitational wave of frequency $\wg$ either directly couples to the electromagnetic fields (left inset) or indirectly by exciting the mechanical vibrational modes at frequencies $\w_p$ (right inset), thereby sourcing electromagnetic power at $\w_0 \pm \wg$. Thus, the signal mode $E_1$ at frequency $\w_1$ is resonantly excited if $\wg \simeq |\w_1 - \w_0|$, which is read out by a directional coupler centered around $\w_1$. The mode profiles of the mechanical vibrations (as indicated by the solid boundary of the cells) and the electromagnetic modes (orange and blue lines) are shown for an optimal configuration. A scan across various gravitational wave frequencies amounts to tuning the electromagnetic frequency difference $\w_1-\w_0$, which can be performed by, e.g., varying the diameter of the central aperture connecting the two cells.}
 \label{fig:Cartoon}
\end{figure}

The experimental setup is as follows. A superconducting radio-frequency (SRF) cavity is prepared with two EM resonant modes at frequencies $\w_0$ and $\w_1$. The first of these we call the \textit{pump mode}, which is to be loaded with EM energy. A GW of frequency $\wg \ll \w_0 , \w_1 \sim 1 \ \GHz$ interacts with the pump mode and sources EM power at $\w_0 \pm \wg \gg \wg$, thereby constituting a frequency-conversion or ``heterodyne" process. The mode at $\w_1$ is initially empty and can be tuned such that the GW frequency $\wg$ matches the mode splitting, $\wg \simeq |\w_1 - \w_0|$, allowing for resonant transfer of EM energy from the pump mode into this \textit{signal mode}. As depicted schematically in \Fig{Cartoon}, two coexisting signal mechanisms can induce this transition: (1) the direct coupling of gravity to the EM energy in the pump mode through the inverse Gertsenshtein effect~\cite{gertsenshtein1962wave,Zeldovich:1972mn}, and (2) the coupling of gravity to the mechanical body of the cavity, which induces EM mode-mixing. As originally noted in Ref.~\cite{Pegoraro:1978gv} and discussed in further detail here, the latter signal is parametrically enhanced compared to the former for $\wg \ll \w_0$. In this work, we show for the first time the potential reach of a broadband setup, where the frequency splitting between EM modes is held fixed. We find that operating the MAGO detector in this manner has the potential to probe orders of magnitude of new parameter space in the 10~kHz to GHz range. The large quality factors $Q\sim 10^{11}$ of SRF cavities allow them to act as efficient converters of mechanical to EM energy and operate with much smaller readout noise than the mechanical-EM transducers employed in modern Weber bar experiments~\cite{Weber:1960zz,Weber:1966zz,Vinante:2006uk,Astone:2002ej,Astone:2002ej}. In this sense, the optimal setup described here functions as a Weber bar with significantly reduced EM noise, resulting in increased sensitivity to GW frequencies that are outside the bandwidth of the mechanical resonance. This is discussed in more detail in \Sec{comparison}. As a result, even for fixed EM frequency splittings, in which case most GW frequencies can only excite the signal off resonance, the reduced EM noise allows this setup to potentially operate as an exquisite broadband detector of high-frequency GWs. In this case, such a search has the added benefit of being sensitive to transient signals that would otherwise be missed by a scanning experiment. For the analysis in this paper, we will consider spherical-cell SRF cavities (such as those employed in the MAGO prototype), since their enhanced symmetry allows greater coverage of the GW sky as well as the availability of analytic results for the various mode profiles.\footnote{Spherical Weber bars have been studied in, e.g., Refs.~\cite{Coccia:1995yi,Lobo:1995sc,Zhou:1995en,Stevenson:1996rw,Gottardi:2006gn,Gottardi:2007zn}.} However, this setup can be applied to any cavity geometry, including the elliptical cavities currently used for state-of-the-art SRF systems. 

Compared to previous work, we introduce three new results: 1) we compute a new source of signal from the direct coupling between the GW and the EM energy in the cavity, 2) we discuss the sensitivity of a broadband operation of the experimental apparatus, where the parameters of the cavity are not resonantly tuned to the GW frequency, and 3) we analytically determine the GW-mechanical and mechanical-EM coupling for spherical cavities, allowing us to estimate the sensitivity as a function of the GW's polarization and direction of propagation. The outline of this paper is as follows. In \Sec{signals}, we discuss signals arising from either the GW-mechanical or GW-EM coupling. In \Sec{modes}, we discuss additional details regarding mechanical and EM mode couplings. In \Sec{noise}, we summarize the main contributions to noise. We use these results in \Sec{sensitivity} to estimate the projected sensitivity of a scanning or broadband setup, and we compare to other experiments such as modern-day Weber bars and lumped-element circuits in \Sec{comparison}. Finally, we conclude and discuss future directions in \Sec{outlook}. Additional technical details regarding our signal and noise estimates are provided in a series of appendices.

\section{Signals}
\label{sec:signals}

We now describe in detail the two classes of aforementioned signals, arising from a GW directly interacting with either the photons or mechanical body of the loaded SRF cavity (see the insets of \Fig{Cartoon}). To leading order in the GW strain $h$, these signals are distinguishable, as each only affects the other at $\order{h^2}$. Therefore, we will treat them separately henceforth. While these two signals are on equal footing conceptually, in practice the mechanical signal is overwhelmingly larger and will determine our analysis and detection strategy. 
An interesting aspect is that, as we will see, the mechanical signal dominates over the  EM one even for GWs that do not resonantly excite a mechanical mode, such that additional tuning of the mechanical resonances is not necessary to achieve sensitivity to strains smaller than $10^{-20}$ across a wide range of frequencies. 

Before providing a detailed derivation of these signals, let us first briefly comment on why the mechanical signal dominates. Heuristically, this is due to the fact that the mechanical vibrations of the cavity are much less ``stiff" than the EM resonances, arising from the small speed of sound in typical solids compared to the speed of light. As a result, the internal binding forces of the material do not hamper the GW's ability to mechanically distort the system, even for wavelengths much larger than the size of the cavity. Let us now show this in more detail. Throughout this work, we take the GW frequency to be much smaller than the typical EM resonant frequency of the cavity, $\wg \ll \w_0 \sim 1 \ \GHz$, which is true in all of the parameter space of interest. As derived in, e.g., Refs.~\cite{Berlin:2021txa,Domcke:2022rgu}, GWs interact directly with the EM field of the pump mode $E_0 \sim \order{10} \ \text{MV} / \text{m}$ and source an effective current of the form $\jeff \sim (E_0 / \Lcav) \, \hpd \, e^{i (\w_0 \pm \w_g) t}$, where $\hpd$ is the  GW amplitude in the proper detector (PD) frame\footnote{In Ref.~\cite{Berlin:2021txa}, we showed that the proper detector frame is the appropriate one for cavity experiments. We also provided a resummation of the metric in that frame to all orders in $\wg \Lcav$, which was necessary for $\wg \Lcav \sim \order{1}$. Here, since we are instead interested in the case where $\wg \Lcav \ll 1$, we only keep the leading order terms in $\wg \Lcav$.} and $\Lcav \sim \w_0^{-1}$ is both the characteristic length scale of the cavity and the scale over which the EM fields vary by an $\order{1}$ fraction. When $\wg \simeq |\w_1 - \w_0|$, this current resonantly excites the EM field of the signal mode at the level of
\be
\label{eq:EMparametric}
E_\text{sig}^{(\text{EM})} \sim Q_\text{em} \, \hpd \, E_0 \sim Q_\text{em} \, (\wg \, \Lcav)^2 \, \htt \, E_0
~,
\ee
where $Q_\text{em}$ is the EM quality factor of the cavity, and in the second equality we related the GW amplitude in the PD frame to that in the transverse-traceless (TT) frame by $\hpd \sim (\wg \Lcav)^2 \, \htt$. 

In comparison, the mechanically-induced signal stems from the tidal force imparted to the cavity walls by the GW, $F_\text{sig}^{(\text{mech})} \sim (\Mcav / \Lcav) \, \hpd$, where $\Mcav$ is the mass of the cavity. Note that although $F_\text{sig}^{(\text{mech})} / E_\text{sig}^{(\text{EM})}$ is independent of $\wg$, crucially the mechanical signal in this setup does not measure force, but rather displacement, which introduces favorable scaling in the quasistatic limit $\wg \Lcav \ll 1$. In particular, outside the bandwidth of a mechanical resonance, the force imparted by the GW displaces the cavity walls by a fractional amount 
\be
\frac{\Delta x_\text{sig}^{(\text{mech})}}{\Lcav} \sim
\frac{F_\text{sig}^{(\text{mech})} }{ \Mcav \, \Lcav} ~ \frac{1}{ \max{(\wg , c_s / \Lcav)}^2} 
\sim 
\htt \, \min{\Big(1 \, , \frac{\wg \, \Lcav}{c_s} \Big)^2}
~,
\ee
where the ``max" and ``min" quantities incorporate the elastic response of the cavity with a speed of sound $c_s \ll 1$ in the limit that the mechanical frequency is higher or lower than that of the GW. This displacement mixes the two EM modes, resonantly exciting the signal field at the level of $E_\text{sig}^{(\text{mech})} \sim Q_\text{em} \, (\Delta x_\text{sig}^{(\text{mech})} / \Lcav) \, E_0$, yielding 
\be
\label{eq:Mechparametric}
E_\text{sig}^{(\text{mech})} 
\sim Q_\text{em} \, \htt \, E_0 \, \min{\Big(1 \, , \frac{\wg \, \Lcav}{c_s} \Big)^2}
~.
\ee
We see that due to the cavity's small speed of sound $c_s \sim 10^{-6}$, the mechanically-induced signal of \Eq{eq:Mechparametric} dominates over the direct EM signal in \Eq{eq:EMparametric}. In particular, the former is enhanced by $(\GHz/\wg)^2$ for $c_s/\Lcav \sim \kHz \ll \wg \ll 1/\Lcav \sim \GHz$ and $1/c_s^2$ for $\wg \ll \kHz$ for a cavity of size $\Lcav \sim \order{10} \ \cm$. In Secs.~\ref{sec:mechsignal} and \ref{sec:EMsignal}, we substantiate these parametric estimates, incorporating further details regarding the GW-EM, GW-mechanical, and mechanical-EM coupling of the cavity. 

The discussion presented throughout this work is valid in the quasistatic limit, $\wg \ll \Lcav^{-1}\sim 1 \ \GHz$. In particular, we will restrict our analysis to $\wg \leq \w_0 / 10$. For completeness, let us briefly comment on the alternative situation that $\wg \gtrsim \Lcav^{-1}$. In this case, the mechanical signal is no longer parametrically enhanced compared to the direct EM one, as can already be seen by comparing Eqs.~\eqref{eq:EMparametric} and~\eqref{eq:Mechparametric}. As a result, an analogous search at higher frequencies is not likely to yield  sensitivity that is competitive with other proposed techniques, since, e.g., a resonant cavity setup with a much larger applied \emph{static} $B$-field can achieve greater sensitivity for $\wg \sim \Lcav^{-1}$~\cite{Berlin:2021txa}. Furthermore, at the level of the signal analysis, the high-frequency regime requires employing the fully resummed metric in the PD frame~\cite{Berlin:2021txa}. 

\subsection{Scanning vs. Broadband} 

Before discussing the signals at a more detailed level in the sections below, let us briefly comment on the two different experimental approaches discussed throughout this work. In particular, we will investigate the potential sensitivity of an apparatus operated in a manner in which the EM mode splitting $\w_1 - \w_0$ is either tuned to the GW frequency or held fixed, corresponding to a scanning (i.e., resonant) or non-scanning (i.e., broadband) setup, respectively. Although the sensitivity of a scanning search exceeds that of a broadband one on general grounds, there are various reasons to consider the latter. First, designing a cavity that scans across orders of magnitude in frequency space is non-trivial. Second, often underemphasized in the literature, is that it is highly time- and labor-intensive to scan across a wide range of frequencies when searching for highly coherent signals. For instance, although we will assume in this work that it takes a time $t_e = 1\ \text{yr}$ to scan across a single $e$-fold in GW frequency (comparable to many resonant axion dark matter experiments), this does not account for the time required to tune the parameters of the experimental apparatus at each new point in $\w_1 - \w_0 = \wg$. In practice, an experiment that nominally should take a few years to run might in actuality require on the order of a decade to obtain its projected sensitivity. Third, and perhaps most important, is the fact that a scanning experiment may easily miss transient signals whose frequencies evolve on timescales comparable to the short amount of experimental time spent at each frequency step. 

\begin{figure}[t]
    \centering
    \includegraphics[width=0.7\textwidth]{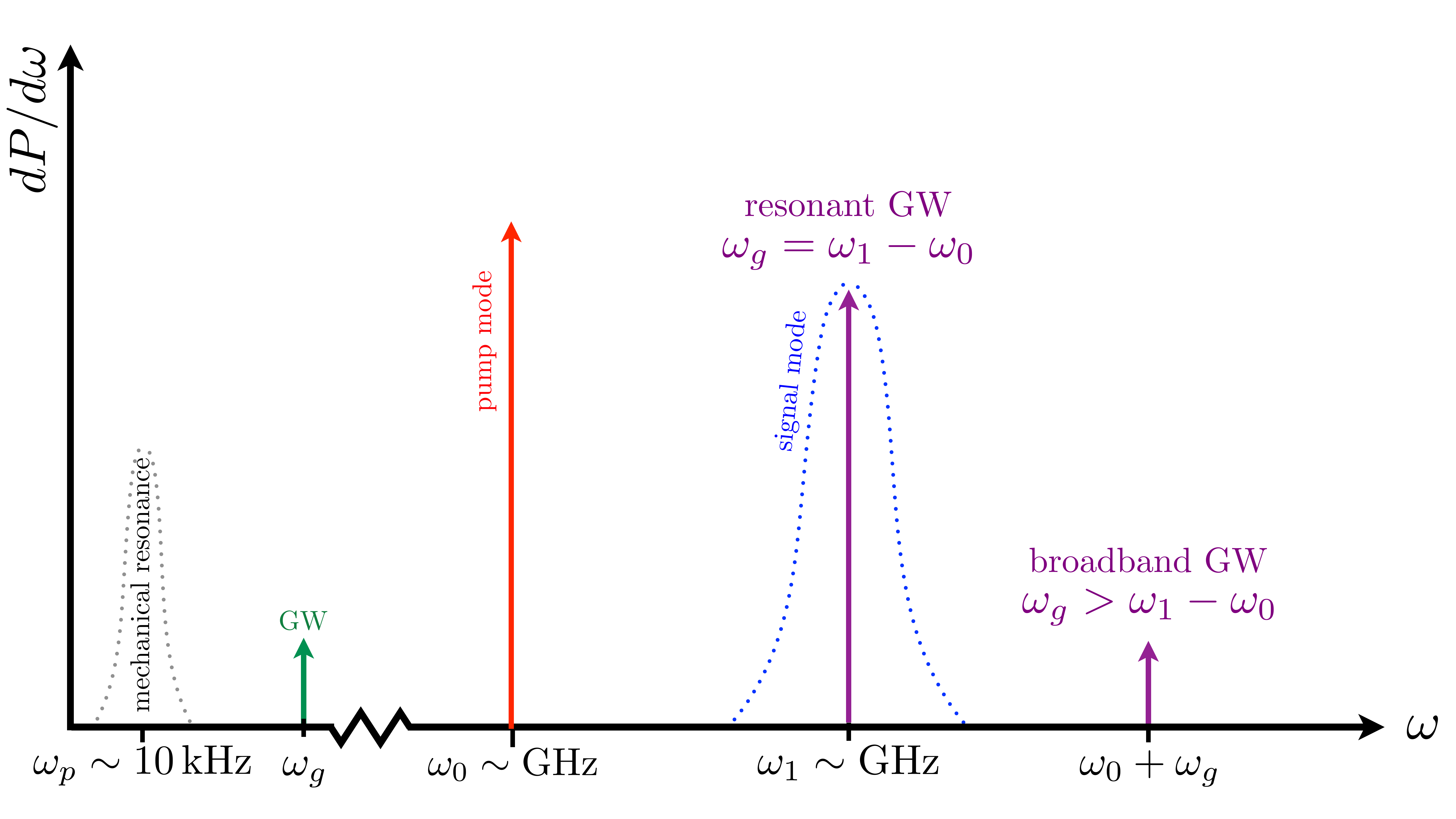}
    \caption{Schematic of the frequency power spectrum for the experimental setup. A gravitational wave with frequency $\wg$ (green) drives a low-lying mechanical mode (dotted black) above its resonant frequency $\w_p$, thereby exciting a small fraction of pump mode photons at $\w_0 \sim 1 \ \text{GHz}$ (red) to signal photons at a frequency $\w_0 + \wg$. In a ``scanning" setup, the electromagnetic mode splitting is fixed to match the gravitational wave frequency, such that such signal photons are within the bandwidth of the electromagnetic signal mode (dotted blue) and are thus resonantly amplified (tall purple arrow). In a ``broadband" setup, the electromagnetic mode splitting is held fixed, and generically the electromagnetic signal at $\w_0 + \wg$ is not resonantly excited (short purple arrow).}
    \label{fig:PSDcartoon}
\end{figure}

A few schematic examples of GW signals are shown in \Fig{PSDcartoon}. A generic high-frequency GW ($\wg \gtrsim 10 \ \text{kHz}$) couples to the cavity by driving a low-lying mechanical mode off resonance. The induced vibrational motion of the cavity causes pump mode photons at  $\w_0 \sim 1 \ \text{GHz}$ to be excited to signal photons at $\w_0 + \wg$. Thus, if the EM modes are tuned such that, e.g., $\w_1 \simeq \w_0 + \wg$, then the signal mode is resonantly excited with a response that is enhanced by the large EM quality factor $Q_\text{em}$ of the cavity. If, on the other hand, the EM mode splitting is held fixed, most GWs will excite the signal mode off resonance with significantly suppressed power. However, since most noise sources are similarly suppressed at frequencies  comparable to the signal frequency, such a broadband setup is still sensitive to small GW strains. In this case, such non-resonant signals are not directly amplified by the large EM quality factor. Nevertheless, $Q_\text{em} \gg 1$ still indirectly aids the sensitivity of such an experiment, as it suppresses the strength of irreducible EM noise. 

\subsection{Mechanical Signal} 
\label{sec:mechsignal}

Gravitational waves interact with the mass of the entire system, inducing a tidal force that shakes and deforms the cavity at the frequency $\wg$. Working in the PD frame, the force-density imparted by the GW in the long-wavelength limit is $f_i \simeq - R_{i0j0} \, x^j \, \rho_\text{cav}$, where $R_{\mu\nu\rho\sigma}$ is the Riemann tensor, $x^j$ the spatial coordinate with respect to the cavity's center of mass, and $\rho_\text{cav}$ the cavity mass density~\cite{Maggiore}. Since $R_{\mu\nu\rho\sigma}$ is invariant under coordinate transformations at $\order{h}$, it is convenient to compute it in the transverse-traceless (TT) frame, such that $R_{i0j0} \simeq - \ddot{h}_{ij}^\text{TT} / 2$.
%}
The resulting displacement of the cavity wall away from its equilibrium position is decomposed as $\U (\xv , t) = u_p (t) \, \U_p (\xv)$, where $\U_p$ are the dimensionless spatial profiles of the $p^\text{th}$ mechanical normal mode (normalized to unity when averaged over the volume $\Vshell$ of a single thin spherical shell)  and $u_p$ the corresponding time-dependent amplitude. 
Also parametrizing the coherent GW as $\htt_{ij} = h_0 \, \htthat_{ij} e^{i \wg t}$, where $h_0$ is the characteristic strain amplitude in the TT frame and $\hat{h}_{ij}^\text{TT} \sim \order{1}$, the equation of motion governing the mechanical coupling of the GW to the cavity is then~\cite{Maggiore}
\be
\label{eq:EOMmech}
\ddot{u}_p + \frac{\w_p}{Q_p} \, \dot{u}_p + \w_p^2 \, u_p \simeq - \frac{1}{2} \, \wg^2 \, \Vcav^{1/3} \, \eta_\text{mech}^g \, h_0 \, e^{i \wg t}
~.
\ee
Above, $\w_p$ is the mechanical resonant frequency, $Q_p$ the mechanical quality factor, $\Vcav$ the volume of a single spherical cavity, and we have defined the dimensionless coefficient $\eta_\text{mech}^g$ that quantifies the GW-mechanical coupling,
\be
\label{eq:GWmechcoupling}
\eta_\text{mech}^g = \frac{\htthat_{ij}}{\Vcav^{1/3} \, \Vshell} \, \int_{\Vshell} \hspace{-0.3cm} d^3 \xv ~ U_p^{* i} \, x^j
~,
\ee
where $\Vshell \ll \Vcav$ is the volume of conducting material in one of the thin spherical shells.

The deformation $u_p (t)$ of the cavity acts as a time-dependent perturbation to the Hamiltonian that has non-zero overlap between the pump and signal modes, thereby transferring power between the two (a process which is resonantly enhanced if $\wg \simeq |\w_1 - \w_0|$). This can be seen at the level of the classical equations of motion. We first decompose the electric field as $\E (\xv , t) = e_i (t) \, \E_i(\xv)$, where $\E_i$ is the spatial profile of the pump ($i=0$) and signal ($i=1$) mode, and $e_i$ the corresponding time-dependent amplitude (and similarly for the $B$-fields).  As derived in Appendices~\ref{app:Cavity} and \ref{app:mechanical-em-overlaps}, the  mechanical perturbation $u_p$ couples to the EM modes $e_{0,1}$ through
\be
\label{eq:EOMem}
\ddot{e}_1 + \frac{\w_1}{Q_1} \, \dot{e}_1 + \w_1^2 \, e_1 \simeq - 2 \, \eta_\text{mech}^\text{EM} \, \w_1^2 \, \Vcav^{-1/3} ~ u_p \, e_0
~,
\ee
where $Q_1$ is the EM quality factor of the signal mode. We have also defined an additional dimensionless coefficient $\eta_\text{mech}^\text{EM}$ that controls the mechanical-EM coupling
\be
\label{eq:etame}
\eta_\text{mech}^\text{EM} = \Vcav^{1/3} ~ \frac{\int_{S_0} d \A \cdot \U_p ~ \big( \E_0 \cdot \E_1^* - \B_0 \cdot \B_1^* \big)}{\int_{\Vcav} \hspace{-0.1cm} d^3 \xv ~ |\E_1|^2}
~,
\ee
where the integral in the numerator is performed over the unperturbed cavity surface $S_0$. From the right-hand side of \Eq{eq:EOMem}, we see that GW-induced vibrations (oscillating at $\wg$) and the pump mode (oscillating at $\w_0$) act as a driving term that can resonantly excite the signal mode if $\w_0 \pm \wg = \w_1$. Furthermore, \Eq{eq:etame} implies that this process is optimized for EM field profiles that are aligned between the two modes. In \Sec{modes} below, we will revisit the requirements to achieve $\eta_\text{mech}^g \, , \, \eta_\text{mech}^\text{EM} \sim \order{1}$.

The equations of motion in Eqs.~(\ref{eq:EOMmech}) and (\ref{eq:EOMem}) can be solved after Fourier transforming to frequency space. As derived in \App{mechPSDs}, the power spectral density (PSD) of the signal power arising from a coherent source of GWs coupling to the $p^\text{th}$ mechanical resonance is given by
\be
\label{eq:sigPSD}
S_\text{sig}^{(\text{mech})} (\w) \simeq %\frac{1}{16} 
\frac{1}{4}
\, \frac{Q_\text{int}}{Q_\text{cpl}} ~ |\eta_\text{mech}^g|^2 \, |\eta_\text{mech}^\text{EM}|^2 ~ P_\text{in} \, h_0^2   ~~ \frac{\w_0^4 ~ S_{\hat{e}_0} (\w - \wg)}{(\w^2 - \w_1^2)^2 + (\w \, \w_1 / Q_1)^2} ~~ \frac{\wg^4}{(\wg^2 - \w_p^2)^2 + (\wg \, \w_p / Q_p)^2}
~,
\ee
where we have adopted the PSD conventions of Refs.~\cite{Berlin:2019ahk,Berlin:2020vrk}. In \Eq{eq:sigPSD}, $Q_\text{int}$ is the \emph{intrinsic} EM quality factor of the cavity and $P_\text{in} = (\w_0 / Q_\text{int}) \int d^3 \xv \, e_0^2 \, |\E_0|^2$ is the input power of the pump mode. The parameter $Q_\text{cpl}$ controls the degree of coupling to the readout, which dictates the loaded quality factor of the signal mode by $Q_1^{-1} = Q_\text{int}^{-1} + Q_\text{cpl}^{-1}$. Hence, a ``critically-coupled" or ``over-coupled" setup corresponds to $Q_\text{cpl} = Q_\text{int}$ or $Q_\text{cpl} \ll Q_\text{int}$, respectively. In \Eq{eq:sigPSD}, we have also introduced the PSD $S_{\hat{e}_0} (\w)$ of the pump mode waveform, such that a monochromatic pump mode corresponds to $\hat{e}_0 = \cos{\w_0 t}$. For a coherent GW, we must account for the spectral spread of the pump mode itself, which is determined by the width $\Delta \w_\text{osc}$ of the external oscillator used to drive the cavity~\cite{Berlin:2020vrk}. Approximating the external oscillator as spectrally flat within a narrow region centered around $\w_0$, we have
\be
S_{\hat{e}_0} (\w) \simeq  \frac{\pi^2}{\Delta \w_\text{osc}} \, \Theta (\Delta \w_\text{osc} / 2 - |\w - \w_0|)
~.
\ee
In our estimates, we assume that this is much narrower than the signal mode bandwidth (but still resolvable within an integration time of $t_\text{int}$) such that $t_\text{int}^{-1} \ll \Delta \w_\text{osc} \ll \w_1 / Q_1$, which is true for commercially available oscillators~\cite{datasheet}. In the case that the GW is resonant with the EM splitting ($\wg \simeq |\w_1 - \w_0|$), the integrated form of \Eq{eq:sigPSD} determines the total signal power to be
\be
\label{eq:mechsigpower}
P_\text{sig}^{(\text{mech})} \simeq %\frac{1}{16} \, 
\frac{1}{4} \, \frac{Q_1^2 \, Q_\text{int}}{Q_\text{cpl}} \, |\eta_\text{mech}^g|^2 \, |\eta_\text{mech}^\text{EM}|^2 ~  P_\text{in} \, h_0^2 \times
\begin{cases}
\frac{\wg^4}{(\wg^2 - \w_p^2)^2} & ,~ |\wg - \w_p| \gg \w_p / Q_p
\\
Q_p^2 & ,~  |\wg - \w_p| \ll \w_p / Q_p
~,
\end{cases}
\ee
where the two cases correspond to whether the GW is off (top) or on (bottom) resonance with the mechanical mode. 
Alternatively, in the case that the GW signal is not resonant with the signal mode (e.g., if $\wg \gg |\w_1 - \w_0| \gg \w_0 / Q_1$), then the signal power is suppressed by a factor of $\sim \w_0^2 / (Q_1  \, \wg)^2 \ll 1$ compared to \Eq{eq:mechsigpower}. From \Eq{eq:mechsigpower} we can glean several features. First, note that when the GW is resonant with the mechanical mode $\wg \simeq \w_p$, the signal power is independent of $\wg$ and enhanced by $Q_p$. For $\wg \gg \w_p$, the signal remains independent of $\wg$, but is no longer enhanced by $Q_p$. Finally, if $\wg \ll \w_p$ for all mechanical modes, the signal power decouples as $(\wg / \w_p)^4$, which  is expected on general grounds, since in the $\wg\to 0$ limit we recover flat space. 

\begin{figure}[t]
\centering
\includegraphics[width=0.5\textwidth]{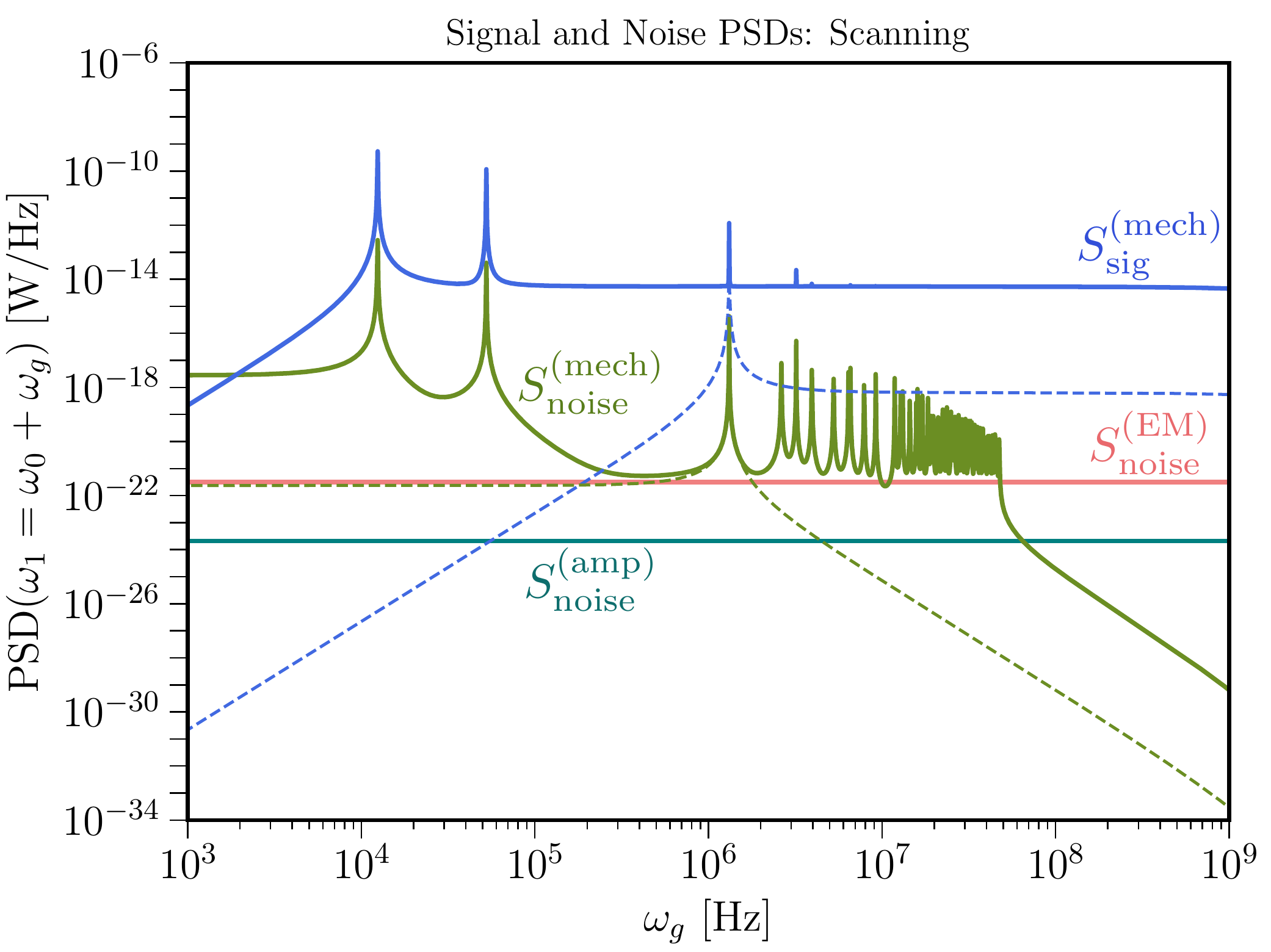}~\includegraphics[width=0.5\textwidth]{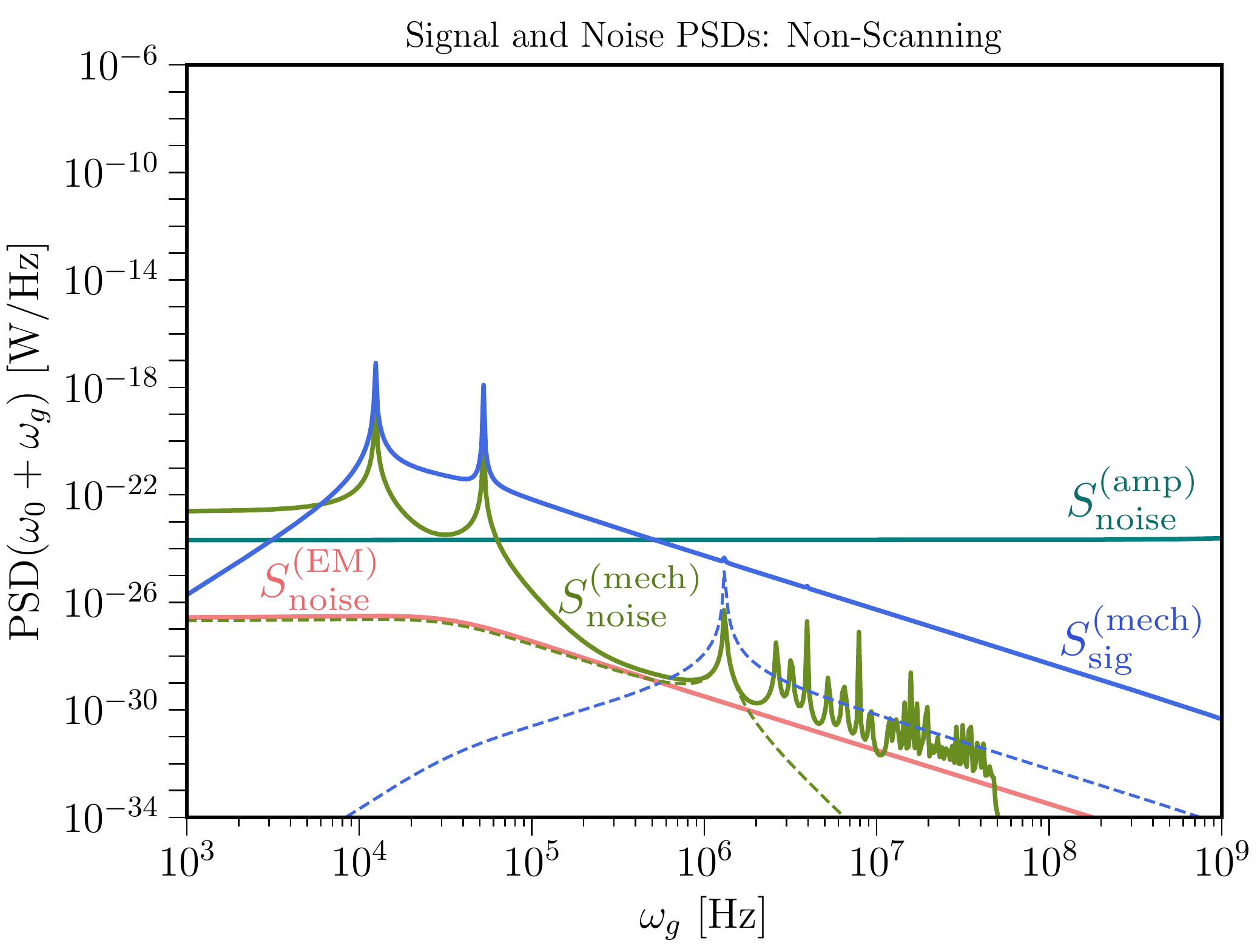}
\caption{Power spectral densities (PSDs) evaluated at $\w_0 + \wg$ for the total mechanical signal (blue) and noise arising from thermal occupation of the EM modes (pink), thermal occupation of the mechanical modes (green), and a quantum-limited amplifier (cyan). The characteristic strain of the coherent gravitational wave is set to $h_0 = 10^{-20}$. 
The contribution of a single excited mechanical mode at $\sim 1 \ \text{MHz}$ to the signal (dashed blue) and noise (dashed green) is also shown. See \Sec{noise} for further details regarding noise. The parameters of the cavity are the same as shown later in \Fig{Reach}.
\textbf{Left}: PSDs for a scanning setup, in which the EM mode splitting is fixed to the GW frequency, $\w_1 - \w_0 = \wg$ and the quality factor of the signal mode is $Q_1 = 10^{10}$.
\textbf{Right}: PSDs for a non-scanning setup, in which $Q_1 = 10^5$ and the EM mode splitting is fixed to the lowest-lying spin-2 mechanical resonance,   $\w_1 - \w_0 = \min \w_p \sim 10 \ \text{kHz}$ with mechanical quality factor $Q_p = 10^6$.
}
\label{fig:PSDs}
\end{figure}

Special care needs to be taken when considering the GW coupling to the mechanical modes of the cavity, which we model as two independent hollow spheres (numerical simulations have shown that this is a good approximation for frequencies above $\sim 1 \ \kHz$~\cite{Ballantini:2005am}). First, as discussed in \App{GWmechOverlap}, GWs couple only to the so-called ``spheroidal'' modes (i.e., vibrations related to changes in volume, as opposed to fixed-volume ``toroidal" modes arising from rotation/shear) of the system~\cite{Maggiore}. Not only do GWs solely couple to such modes, but from angular momentum selection rules it can be shown that only $l=2$ modes can be excited to leading order in $\wg \Vcav^{1/3} \ll 1$~\cite{Maggiore}. The mechanical response of the cavity is described by a forest of such spin-2 resonances. However, not all resonances contribute equally to the response. To simplify the signal analysis, it is therefore helpful to identify the mechanical mode that leads to the largest signal, as a function of the GW frequency. We find that, independent of $\wg$, the signal is in fact dominated by the lowest-lying $\ell = 2$ mechanical resonance with frequency $\min \w_p \sim 10 \ \kHz$. This is simple to see for $\wg \ll 10 \ \kHz$, since in this case $P_\text{sig} \propto (\wg / \w_p)^4$, as discussed above. For higher frequency GWs, we first note that the GW-mechanical coupling decreases as a function of the number of nodes that the mechanical mode possesses along the radial direction within the cavity walls, since the GW is approximately spatially uniform for $\wg \lesssim \w_0$. As a result, we find in \Sec{modes} that beyond the first few mechanical resonances, the scaling is well-approximated by $\eta_\text{mech}^g \propto 1/\w_p^2$ (this scaling may change significantly for different cavity geometries). Furthermore, the mechanical quality factor is expected to decrease with frequency as $Q_p \propto 1/\w_p$~\cite{Zener:1937zz}. As a result, aside from the special case where $\wg \simeq \w_p$ for the first few mechanical modes, it is the off-resonance response of the lowest-lying spin-2 mode that dominates the signal at most GW frequencies. This is shown in \Fig{PSDs}, where the total signal PSD (shown as the solid blue line) is given by the sum over the contributions of the individual mechanical resonances and is well-approximated by just the response of the lowest-lying mode. In particular, the individual contribution of a higher mechanical resonance is plotted as dotted blue, which is shown to contribute a subdominant fraction of the total signal power. 

\subsection{Electromagnetic Signal}
\label{sec:EMsignal}

Here, we briefly describe the signal arising from the direct interaction between the GW and the EM modes of the cavity. As explained above, this is parametrically suppressed compared to the mechanically-induced signal in the $\wg \Vcav^{1/3} \ll 1$ limit. However, we note that the sensitivity of an experiment optimized for this signal could be enhanced, in principle, if vibrational noise was significantly reduced using, e.g., active feedback. Although, as we describe below in \Sec{sensitivity}, achieving sensitivity comparable to the mechanically-induced signal does not seem feasible, we describe the direct EM signal here for the sake of completeness.

The direct interaction of the GW with the pump mode sources an effective current, which in the PD frame is given by~\cite{Berlin:2021txa,Domcke:2022rgu}
\be
\label{eq:jeff}
\jeff^\mu = \partial_\nu \bigg[ \, \hpd_{\alpha \beta} \, \bigg( \, \frac{1}{2} \, \eta^{\alpha \beta} \, F^{\mu \nu} + \eta^{\beta \nu} \, F^{\alpha \mu} - \eta^{\beta \mu} \,  F^{\alpha \nu} \bigg) \bigg]  
\equiv \big( \wg \Vcav^{1/3} \big)^2 \, h_0 \, \w_0 \, \langle E_0 \rangle \, \hat{j \, }^\mu (\xv) \, e^{i (\w_0 \pm \w_g) t}
~,
\ee
where in the above expression $\langle E_0 \rangle$ is taken to be the volume-averaged amplitude of the pump field, $\hat{j \, } (\xv)$ is a dimensionless vector that accounts for the EM mode profiles and the GW polarization, and the factor of $(\wg \Vcav^{1/3})^2$ arises from relating the metric in the PD frame to that in the TT frame $\hpd \sim (\wg \Vcav^{1/3})^2 \, h_0$. When $\w_g \simeq |\w_1 - \w_0|$, this resonantly excites the signal mode, leading to a total integrated signal power of~\cite{Berlin:2019ahk, Berlin:2020vrk, Berlin:2021txa}
\be
\label{eq:EMsigpower}
P_\text{sig}^{(\text{EM})} \simeq \frac{1}{2} \, \frac{Q_1^2 \, Q_\text{int}}{Q_\text{cpl}} \, |\eta_\text{EM}^g|^2 \,  P_\text{in}
\, h_0^2 \times \big( \wg \Vcav^{1/3} \big)^4
~,
\ee
where we have defined the dimensionless GW-EM coupling
\be
\eta_\text{EM}^g = \frac{\int d^3 \xv ~ \E_1^* \cdot \hat{\jv \, }}{\left( \Vcav 
 \, \int d^3 \xv ~ |\E_1|^2\right)^{1/2}}
~.
\ee
Note that compared to the mechanically-sourced signal in \Eq{eq:mechsigpower}, the direct EM signal in \Eq{eq:EMsigpower} is suppressed by a factor of $\max(\wg, \w_p)^4 \, \Vcav^{4/3}$. This confirms the intuition laid out in the discussion near Eqs.~\eqref{eq:EMparametric} and \eqref{eq:Mechparametric} and justifies our focus on the mechanical signal in this work. It is instructive to compare \Eq{eq:EMsigpower} to the signal power attainable in a setup employing static EM fields. For instance, Ref.~\cite{Domcke:2022rgu} recently proposed a modified version of the ``DMRadio" axion dark matter experiment, consisting of a static $B$-field applied to an LC circuit (whose resonant frequency is matched to $\w_g$). As shown in Ref.~\cite{Domcke:2022rgu}, a distinct readout specialized to the GW-generated EM field (in the form of a ``figure-8" loop pickup) enables the same $\wg$-scaling as shown in \Eq{eq:EMsigpower}.

\section{Mechanical and Electromagnetic Mode Coupling}
\label{sec:modes}

In this section, we highlight key requirements of the cavity design, focusing on optimization of the GW-mechanical and mechanical-EM couplings, as defined in Eqs.~(\ref{eq:GWmechcoupling}) and (\ref{eq:etame}), respectively. An optimal design will enable $\order{1}$ values for both couplings, as well as EM modes with a tunable frequency splitting matched to the GW frequency. We show below that a coupled-cavity setup with judiciously chosen EM polarizations meets these requirements, enabling sensitivity across a broad range of GW frequencies (see \Sec{sensitivity}). A more detailed treatment is given in Appendices~\ref{app:GWmechOverlap} and \ref{app:mechanical-em-overlaps}.

A single spherical cavity with inner radius $a \sim 10 \ \text{cm}$ possesses EM modes separated by a characteristic frequency splitting of $\sim 1 \ \text{GHz}$ and low-lying mechanical modes with frequency $\sim 10 \ \kHz$. To probe sub-GHz GWs, we envisage employing a system similar to the existing MAGO prototype~\cite{Ballantini:2003nt,Ballantini:2005am}, consisting of two nearly-identical spherical cells, electromagnetically coupled via a small circular aperture, as shown in \Fig{Cartoon}. For a narrow aperture, the pump and signal modes of the coupled system are well approximated as consisting of the symmetric (orange lines in \Fig{Cartoon}) and antisymmetric (blue lines in \Fig{Cartoon}) linear combination of single-sphere modes. We therefore denote the pump $\E_0$ and signal $\E_1$ modes of the total cavity  as tensor products of single-sphere modes $\tilde{\E}$ of the left and right cell. In particular, we take the pump and signal modes to have identical field configurations in the left cell, but opposite field profiles in the right cell, such that $\E_0 = \tilde{\E} \, \otimes \, \mathcal{R} \tilde{\E}$ and $\E_1 = \tilde{\E} \, \otimes \, ( - \mathcal{R} \tilde{\E})$ (and similarly for the $B$-fields), where $\mathcal{R}$ is a rotation matrix acting on the right-cell mode, which accounts for the relative spatial orientation between the two spheres.\footnote{The spherical symmetry is broken by the aperture joining the two cavities and hence their relative orientation is physical.} For an aperture of radius $d$ joining two spherical cells of radius $a$, the induced frequency splitting between the two modes is roughly $\Delta \w_\text{EM} \sim (d/a)^3 \, \w_0 \ll \w_0$, where $\w_0 \sim 10 \ \text{GHz}$ is the resonant frequency in the absence of the coupling~\cite{PhysRev.66.163}. Thus, an aperture with a mechanically-tunable radius spanning $\sim 1 \ \text{mm} - \text{few cm}$ induces a splitting of $\sim 1 \ \text{kHz}- 100 \ \text{MHz}$. In our calculations, we ignore collective mechanical oscillations and take the mechanical modes to be those of uncoupled spherical shells.  In Ref.~\cite{Ballantini:2005am}, this was found to be a good approximation for $\wg \gtrsim 10 \ \kHz$. 

\subsection{GW-Mechanical Coupling-Coefficient}
Here we discuss the GW-mechanical coupling $\eta_\text{mech}^g$ of \Eq{eq:GWmechcoupling} for the setup described above. Since the GW directly excites the mechanical mode of the cavity, we take the GW's direction of propagation and the mechanical profile $\Uv_p$ to be oriented along the same axis. In this case, we can calculate $\eta^{g}_{\text{mech}}$ for any mechanical mode $p$, as specified by the radial index $n \geq 1$ and two spherical harmonic indices $\ell, m$ defined with respect to the GW axis. The frequencies of these modes are degenerate in the inclination index $m$ and increase with increasing $n$ and $\ell$. As discussed in \App{GWmechOverlap}, GWs only couple to spheroidal $\ell = 2$ mechanical modes. Restricting our analysis to this subset, we find that the optimal coupling is achieved for the $(\ell, n, m) = 212$ mode, which has a resonant frequency of $\w_p \sim 12 \ \kHz$ for a spherical niobium cavity with inner radius $a = 20 \ \text{cm}$ and thickness  $5 \ \text{mm}$. In particular, a numerical determination of this GW-mechanical coupling yields $\eta^{g}_{\text{mech}}\simeq 0.35$. 

An $\order{1}$ GW-mechanical coupling is achieved for only the few lowest-lying spin-2 mechanical modes. This is due to the fact that for $10 \ \kHz \ll \wg \ll 1 \ \GHz$, the GW is approximately uniform throughout the entire cavity volume, whereas the mechanical mode profile is not (see the functional form of $\Uv_p$ in \App{mechanical-modes}), which suppresses the overlap volume integral in \Eq{eq:GWmechcoupling}. In particular, for large radial index $n \geq 3$, the increasing number of radial nodes in $\Uv_p$ gives rise to a coupling that scales as $\eta^{g}_{\text{mech}} \propto 1/ \w_p^2$. In this work, we take the resonant frequencies of the mechanical modes to be fixed. Regardless, as discussed above in \Sec{mechsignal} and shown explicitly in \Sec{sensitivity} below, this setup still has powerful sensitivity to GWs  well outside the bandwidth of the lowest-lying mechanical resonance.

\subsection{Mechanical-EM Coupling-Coefficient}

In this section, we briefly discuss the EM-mechanical coupling $\eta^{\text{EM}}_{\text{mech}}$ of \Eq{eq:etame}. Further details are provided in \App{mechanical-em-overlaps}. As shown below, for an optimal choice of the EM modes, $\eta^{\text{EM}}_{\text{mech}} \sim \order{1}$ is obtained for a significant fraction of GW propagation directions and polarizations. Approximating the left and right cells as spheres of comparable size, \Eq{eq:etame} is given by the sum of individual surface integrals,
\be
\label{eq:etame-coupled}
\eta_\text{mech}^\text{EM} = \frac{\Vcav^{1/3}}{\int d^3 \xv ~ |\tilde{\E}|^2} ~ \bigg[ \int_{S_{0L}} \hspace{-0.2cm} d \A \cdot \U_p ~ \Big( |\tilde{\E}|^2 - |\tilde{\B}|^2 \Big) - \int_{S_{0R}} \hspace{-0.2cm} d \A \cdot \U_p ~ \Big( |\mathcal{R} \tilde{\E}|^2 - |\mathcal{R} \tilde{\B}|^2 \Big) \bigg]
~,
\ee
where $S_{0L}$ and $S_{0R}$ are the unperturbed surfaces of the left and right cavity, respectively. Since the GW wavelength is much larger than the size of the experimental setup, the polarization of $\U_p$ is the same for both cells. Furthermore, since the $\ell = m = 2$ mechanical mode $\U_p$ is odd under a rotation of $\pi/2$, the two integrands in \Eq{eq:etame-coupled} are maximized and add constructively if: 1) the normal component of the mechanical mode and magnitude of the EM mode possess the same symmetry axis and angular structure, and 2) the relative EM orientation $\mathcal{R}$ is given by a rotation of $\pi/2$ along the GW's direction of propagation. An example of such an arrangement is shown in \Fig{Cartoon}. In \App{mechanical-em-overlaps}, we confirm this by numerically evaluating $\eta_\text{mech}^\text{EM}$ for various relative polarizations of the $\text{TE}_{112}$ EM mode in either cell.

\begin{figure}
    \centering
    \includegraphics[width=0.6\textwidth]{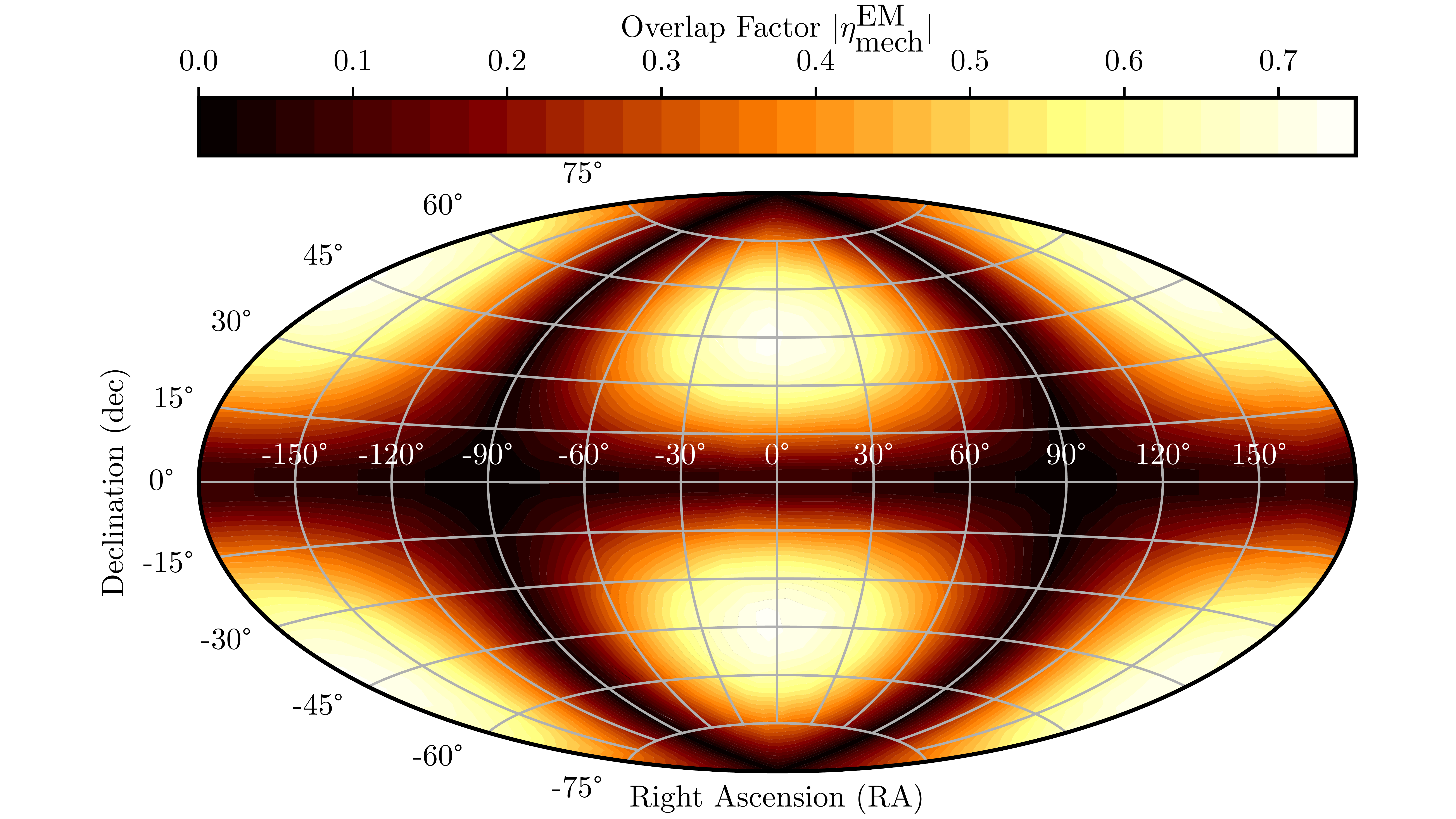}
    \caption{Sky map of the mechanical-electromagnetic coupling $\eta_\text{mech}^\text{EM}$ as a function of the direction of propagation of a plus-polarized GW, where the coordinate and polarization conventions are given in \App{mechanical-em-overlaps}. The map is shown in terms of equatorial coordinates of the GW source, such that the polarization of the excited mechanical mode aligns with the propagation direction of the GW. The pump and signal electromagnetic modes correspond to combinations of $\text{TE}_{112}$ single-sphere modes that are symmetric or antisymmetric across the two cells with a relative offset in polarization by a rotation of $\pi/2$ (refer to the discussion in \Sec{modes} for further details). The mechanical mode of each cell is taken to be the $\ell=2$, $n=1$, $m=2$ spheroidal mode with a polarization fixed by the direction of the GW source. For this choice of mechanical mode, the GW-mechanical coupling is $\eta^{g}_\text{mech} \simeq 0.35$. The maxima, at which $\eta_\text{mech}^\text{EM} \sim \order{1}$, correspond to points where the direction of the incoming GW is aligned with one of the electromagnetic axes.}
    \label{fig:hi-res-couplings-main-body}
\end{figure}

We calculate the mechanical-EM coupling $\eta_\text{mech}^\text{EM}$ as a function of the GW's direction of propagation on the sky for an optimal selection of cavity modes (i.e., for $\Uv_p$ and $\tilde{\E} , \tilde{\B}$ consisting of the $212$ mechanical mode and the $\text{TE}_{112}$ EM mode, respectively), fixing the left/right relative EM orientation to be a rotation of $\pi/2$. This is displayed in \Fig{hi-res-couplings-main-body}, which shows that $\eta_\text{mech}^\text{EM} \sim \order{1}$ across a large fraction of the sky and that the coupling is maximized when the GW's propagation direction is aligned with one of the EM axes. We also note that in contrast to the GW-mechanical coupling of the previous subsection, $\eta_\text{mech}^\text{EM} \sim \order{1}$ for many of the higher excited mechanical resonances as well, since the EM modes only couple to $\Uv_p$ across the cavity surface, as opposed to the volume. For this setup, the normal component of the mechanical mode scales as $\U_p \cdot \hat{\n} \propto \sin^2{\theta} \cos{2 \phi}$, whereas the EM modes evaluated along the surface are $|\tilde{\E}|^2 = 0$ and  $|\tilde{\B}|^2 \propto \cos^2{\theta} \, \cos^2{\phi} + \sin^2{\phi}$. Hence, both the mechanical and EM factors of \Eq{eq:etame-coupled} are quadrupolar and have non-zero overlap when integrated over the cavity surface.

\section{Noise Sources} 
\label{sec:noise}

In this section, we discuss the importance of various noise sources. Our results are summarized in \Fig{PSDs}, which shows noise PSDs for various experimental configurations. The most dominant source of noise in this setup is expected to arise from external vibrations that couple to the cavity and induce a resonant transfer of EM energy, analogous to the mechanical signal of \Sec{mechsignal}. Generalizing \Eq{eq:EOMmech}, the mechanical response of the cavity to an external force is given by 
\be
\label{eq:force1}
\ddot{u}_p + \frac{\w_p}{Q_p} \, \dot{u}_p + \w_p^2 \, u_p = F_p / \Mcav
~.
\ee
Above, $F_p$ is the force projected on the $p^\text{th}$ mechanical mode, which is related to the external force density $\f$ by
\be
F_p = \int_{\Vshell} \hspace{-0.3cm} d^3 \xv ~ \f \cdot \U_p 
~.
\ee
As in \Sec{mechsignal}, we Fourier transform Eqs.~(\ref{eq:EOMem}) and (\ref{eq:force1}) to determine the noise power arising from mechanical vibrations,
\be
\label{eq:mechnoisePSD}
S_\text{noise}^{(\text{mech})} (\w) \simeq %\frac{1}{4} 
 \frac{Q_\text{int}}{Q_\text{cpl}} \,  |\eta_\text{mech}^\text{EM}|^2 \, P_\text{in} ~ \frac{\w_0^4}{(\w^2 - \w_1^2)^2 + (\w \, \w_1 / Q_1)^2} 
~ \, 
\frac{S_{F_p}  (\w - \w_0) \, \Mcav^{-2} \, \Vcav^{-2/3}}{\big( (\w - \w_0)^2 - \w_p^2 \big)^2 + \big( (\w - \w_0) \, \w_p / Q_p \big)^2}
~.
\ee
Appearing in \Eq{eq:mechnoisePSD} is the same mechanical-EM overlap factor, $\eta_\text{mech}^\text{EM}$, as discussed for the GW signal. However, unlike the signal, for which the lowest-lying spin-2 mechanical resonance dominates, \emph{many} resonances will be relevant for mechanical noise. This is due to the fact that, unlike sub-GHz GWs, external source of vibrations need not be spatially uniform across the cavity and thus may have $\order{1}$ spatial overlap with higher mechanical modes.

To estimate the noise power in \Eq{eq:mechnoisePSD}, we see that we need to determine the PSD of the force $S_{F_p} (\w)$, which has reducible contributions from, e.g., seismic noise and the cryogenic system, as well as irreducible thermal fluctuations from the cavity itself. Measurements of SRF cavity microphonics at Fermilab have detected the presence of vibrations at an RMS value of $\langle u_p^2 \rangle^{1/2} \sim 0.1 \ \text{nm}$~\cite{DarkSRF,Pischalnikov:2019iyu}, though no specific attempts were made to improve this because it was sufficient for the particular design goal of the cavity. We use this measurement to determine the force responsible for such vibrations, by noting that the maximal displacement should arise from exciting the lowest mechanical resonance at a frequency of $\min{\w_p} \sim 10 \ \kHz$~\cite{Berlin:2019ahk,Berlin:2020vrk}, 
\be
\label{eq:fermilabforce}
S_{F_p} (\min{\w_p}) \simeq 4 \pi \, \Mcav^2 \, \min(\w_p)^3 \, \langle u_p^2 \rangle \, / \, Q_p 
\sim 10^{-11} \ \text{N}^2 \ \Hz^{-1} \times \bigg( \frac{\Mcav}{10 \ \text{kg}} \bigg)^2 \bigg( \frac{\min{\w_p}}{10 \ \kHz} \bigg)^3 \bigg( \frac{\langle u_p^2 \rangle^{1/2}}{0.1 \ \text{nm}} \bigg)^2 \bigg( \frac{10^6}{Q_p} \bigg)
~.
\ee
Note that \Eq{eq:fermilabforce} defines the size of the force PSD \emph{only} at a frequency of $\min{\w_p} \sim 10 \ \kHz$.\footnote{This is not in contradiction with the previous statement that many mechanical resonances are important in determining the mechanical noise. The total displacement of the cavity walls is dominated by the lowest resonance, but the transition of power between EM modes at a given (higher) frequency can be dominated by higher resonant modes.} We could not find direct measurements of vibrational noise for much larger frequencies, so we perform a conservative extrapolation of this estimate in order to determine the force power for $\w \gg 10 \ \text{kHz}$.
High-$Q$ mechanical resonators operating at MHz frequencies and above have been demonstrated to reach the thermal noise floor~\cite{Goryachev:2014nna}, but these results cannot be easily incorporated into our estimates since the cryogenic environment is qualitatively different than the liquid helium cooling system employed for SRF cavities. Regardless, one can estimate the frequency dependence of $S_{F_p} (\w>\min \w_p)$ from first principles. It is possible to show that most sources of vibrational noise fall off at high frequencies at least as rapidly as $S_{F_p} (\w) \sim 1/\w^2$~\cite{Saulson:1984yg}. In the absence of a direct experimental confirmation, we conservatively take this noise to fall off linearly in the frequency, i.e., $S_{F_p} (\w > \min{\w_p}) \sim S_{F_p} (\min{\w_p}) \times (\min{\w_p} / \w)$, which is a scaling that is much slower than any source of environmental vibrations that we could identify. Noise with a PSD proportional to $1/\w$ exists in a variety of electronic systems (see, e.g., Refs.~\cite{rubiola:hal-00344308,milotti20021} for a review). 

Even for a cavity that is perfectly isolated from its surrounding environment, an irreducible source of vibrations emerges from the temperature of the cavity itself. This can be quickly derived by noting that the energy stored in mechanical vibrations is related to the cavity temperature $T$ via the equipartition theorem, such that $M_\text{cav} \, \w_p^2 \, \langle u_p^2 \rangle / 2 = Q_p \, S_{F_p}^{(\text{th})} (\w_p) / (8 \pi \Mcav \, \w_p) = T / 2$, implying that the irreducible thermal force is given by
\be
\label{eq:thermalforce}
S_{F_p}^{(\text{th})} (\w_p) = 4 \pi \, M_\text{cav} \, \w_p \, T \, / \, Q_p
\sim 10^{-22} \ \text{N}^2 \ \Hz^{-1} \times \bigg( \frac{\Mcav}{10 \ \text{kg}} \bigg) \bigg( \frac{\w_p}{10 \ \kHz} \bigg) \bigg( \frac{T}{2 \ \text{K}} \bigg) \bigg( \frac{10^6}{Q_p} \bigg)
~.
\ee
This same result can be readily obtained from the fluctuation-dissipation theorem~\cite{Kubo_1966}. Note that unlike the mechanical signal in Eqs.~(\ref{eq:sigPSD}) and (\ref{eq:mechsigpower}), which is independent of the cavity mass, a larger mass suppresses mechanical noise since $S_\text{noise}^{(\text{mech})} \propto S_{F_p} (\w_p) \Mcav^{-2}$ scales as $\Mcav^{-2}$ for an external force of fixed strength or as $\Mcav^{-1}$ when intrinsic thermal vibrations dominate. In estimating the projected sensitivity, we will consider both possibilities, corresponding to mechanical noise comparable to existing measurements in \Eq{eq:fermilabforce}, or noise that has been attenuated to the irreducible value in \Eq{eq:thermalforce}. 

Mitigating vibrational noise down to its irreducible thermal value in \Eq{eq:thermalforce} requires a carefully-designed suspension system. This feat is considerably simplified in our setup compared to LIGO due to the higher frequencies that we consider, but might be complicated by the cooling system required to operate a superconducting cavity. The instruments most similar to our own and already in operation include Weber bars, which have succeeded in suppressing vibrational noise down to such levels. For instance, the AURIGA experiment reported a noise power reduction of $10^{-24}$ at a frequency of $\sim 5 \ \kHz$~\cite{Cerdonio:1997hz}.\footnote{Physically, we can envision a system with $N$ cascaded pendulums, such that their small mechanical resonant frequency $\w_\text{pend}$ reduces the displacement of the cavity off-resonance by a factor of $ (\w_\text{pend}/\w)^{2N}$. As an example, for $\w_\text{pend} \sim 1 \ \Hz$, reducing the cavity displacement by $10^{-8}$ at $\w\sim 1 \ \kHz$ (and, hence, the force PSD by $10^{-16}$) requires only two such pendulums.} Furthermore, since external sources of vibrational noise are expected to rapidly fall off with increasing frequency, it is possible that no such suspensions are required. Although not discussed in detail in this work, we expect that attaining the noise value given in \Eq{eq:thermalforce} requires significant suppression of vibrations that emerge from the immediate cryogenic environment. In fact, preliminary design strategies to mechanically isolate the cavity from the liquid helium environment were pursued previously by the MAGO collaboration~\cite{Ballantini:2005am}.

Although mechanical noise dominates throughout most of the parameter space of interest,  thermal occupation of the EM modes and readout noise from the amplifier are relevant at higher frequencies and well outside the resonator bandwidth, respectively. Following Refs.~\cite{Berlin:2019ahk,Berlin:2020vrk}, these are incorporated by their respective PSDs, 
\begin{align}
\label{eq:EMthermalPSD}
S_\text{noise}^{(\text{EM})} (\w) &= 
\frac{Q_1^2}{Q_\text{int} \, Q_\text{cpl}}
~ \frac{4 \pi T \, (\w \, \w_1 / Q_1)^2}{(\w^2 - \w_1^2)^2 + (\w \, \w_1 / Q_1)^2} 
\\
\label{eq:AmpPSD}
S_\text{noise}^{(\text{amp})} (\w) &= \pi \, \hbar \, \w
~.
\end{align}
In \Eq{eq:EMthermalPSD}, we see that thermal noise, being internal to the cavity, is filtered by the cavity response function and has a peak value controlled by the temperature $T$ (which must be smaller than $\sim 2 \ \text{K}$ to maintain superconductivity) and the intrinsic quality factor $Q_\text{int} \gtrsim 10^{10}$. In \Eq{eq:AmpPSD}, we have taken amplifier noise to be approximately spectrally flat within the sensitivity bandwidth (corresponding to the range of frequencies narrowly centered around $\w_0 + \wg \sim 1 \ \text{GHz}$), independent of the cavity resonance, since it arises due to the effective temperature of the external readout system. We have also assumed that the amplifier operates near the so-called standard quantum limit, corresponding to a single photon per unit bandwidth~\cite{PhysRevB.70.245306,PhysRevD.26.1817}, which is the industry standard for currently operating axion dark matter experiments at GHz frequencies~\cite{Brubaker:2016ktl}. We have explicitly included the factor of $\hbar$ here as a reminder that we are considering quantum-limited amplifier noise.

Additional sources of noise may also play a role, such as ``phase noise" of the external oscillator used to drive the cavity in the presence of crosstalk between the pump mode antenna and the signal mode. Mitigation of such phase noise was a major driving factor for the readout design of the MAGO experiment. In particular, the symmetric pump and antisymmetric signal field configurations were driven and readout using so-called ``magic-tee" $180^\circ$ hybrid couplers, which were able to suppress power in such crosstalk by a factor of $\sim 10^{-14}$~\cite{Ballantini:2003nt,Ballantini:2005am}. With this level of crosstalk mitigation, we find that for commercially-available oscillators, phase noise is expected to be much smaller than the mechanically-induced noise described above~\cite{Berlin:2019ahk,Berlin:2020vrk}, but we include it for completeness in our calculations, using the PSD derived in the appendix of Ref.~\cite{Berlin:2020vrk}. We also do not expect significant contributions from superconducting non-linearities; such effects source power at odd harmonics of the pump mode with frequency-spread governed by the narrow bandwidth of the external oscillator~\cite{Sauls:2022sax}, and hence such noise does not have support near the signal frequency $\w_0 + \wg$.

\section{Experimental Parameters and Expected Sensitivity}
\label{sec:sensitivity}

We now apply the formalism developed above to estimate the sensitivity to \emph{coherent} GWs in the kHz$-$GHz frequency range. In doing so, we assume an apparatus of the form similar to the MAGO prototype, consisting of two  high-$Q$ superconducting spherical cavity cells joined by a small circular tunable aperture. As discussed in \Sec{modes}, the diameter of this aperture controls the splitting between the pump and signal fields, which for the MAGO prototype was designed to scan across $4 \ \kHz - 20 \ \kHz$~\cite{Ballantini:2005am}. As our work is primarily a theoretical study, whose goal is to motivate further development of various design strategies, we do not focus on a particular tuning and readout design in our estimates below. Instead, we consider the sensitivity to GWs across a much wider range of frequency splittings. In doing so, we hold various experimental parameters (such as crosstalk rejection) fixed to the values achieved by the MAGO collaboration more than 15 years ago~\cite{Ballantini:2003nt,Ballantini:2005am}. It is conceivable that at the design stage different cavities can be optimized to target different frequency ranges. 

For concreteness, we adopt the following baseline cavity parameters for our projections: a cavity volume $\Vcav = 30 \ \text{L}$ and mass $\Mcav=10\ \text{kg}$, a temperature $T=1.8\ \text{K}$, a characteristic pump-mode EM field strength $E_0 = 30\ \text{MV/m}$, a mechanical quality factor $Q_p = 10^6$ for the lowest-lying mechanical resonance, and an intrinsic EM quality factor $Q_\text{int} = 10^{10}$. The coupling-coefficients  $\eta_{\rm mech}^g$ and  $\eta^{\rm EM}_{\rm mech}$ are computed assuming a MAGO-like geometry (see Secs.~\ref{sec:signals} and \ref{sec:modes}). In particular, we fix $\eta^{g}_{\text{mech}}= 0.35$ and $\eta_\text{mech}^\text{EM}= 1$ in our analysis. Although we find that the direct GW-EM coupling is significantly smaller for such a cavity ($\eta_{\rm EM}^g < \eta_{\rm mech}^g$), we fix $\eta_{\rm EM}^g = \eta_{\rm mech}^g$ in our projections to more meaningfully compare the mechanical and EM signals in setups dedicated for either one. In regards to experimental parameters and noise estimates, we make assumptions different than that previously assumed for MAGO in only two respects: 1) As detailed  in \Sec{noise}, we update the force PSD responsible for vibrational noise, utilizing recent measurements of cavity microphonics at Fermilab. 2) We assume the frequency splitting between the EM modes to be tunable across a much broader range. Here, we do not specify the scanning mechanism, but note that detailed design strategies along this direction are in progress at SLAC in relation to a prototype axion dark matter experiment~\cite{Berlin:2019ahk}. 

\begin{figure}[t]
\centering
\includegraphics[width=0.8\textwidth]{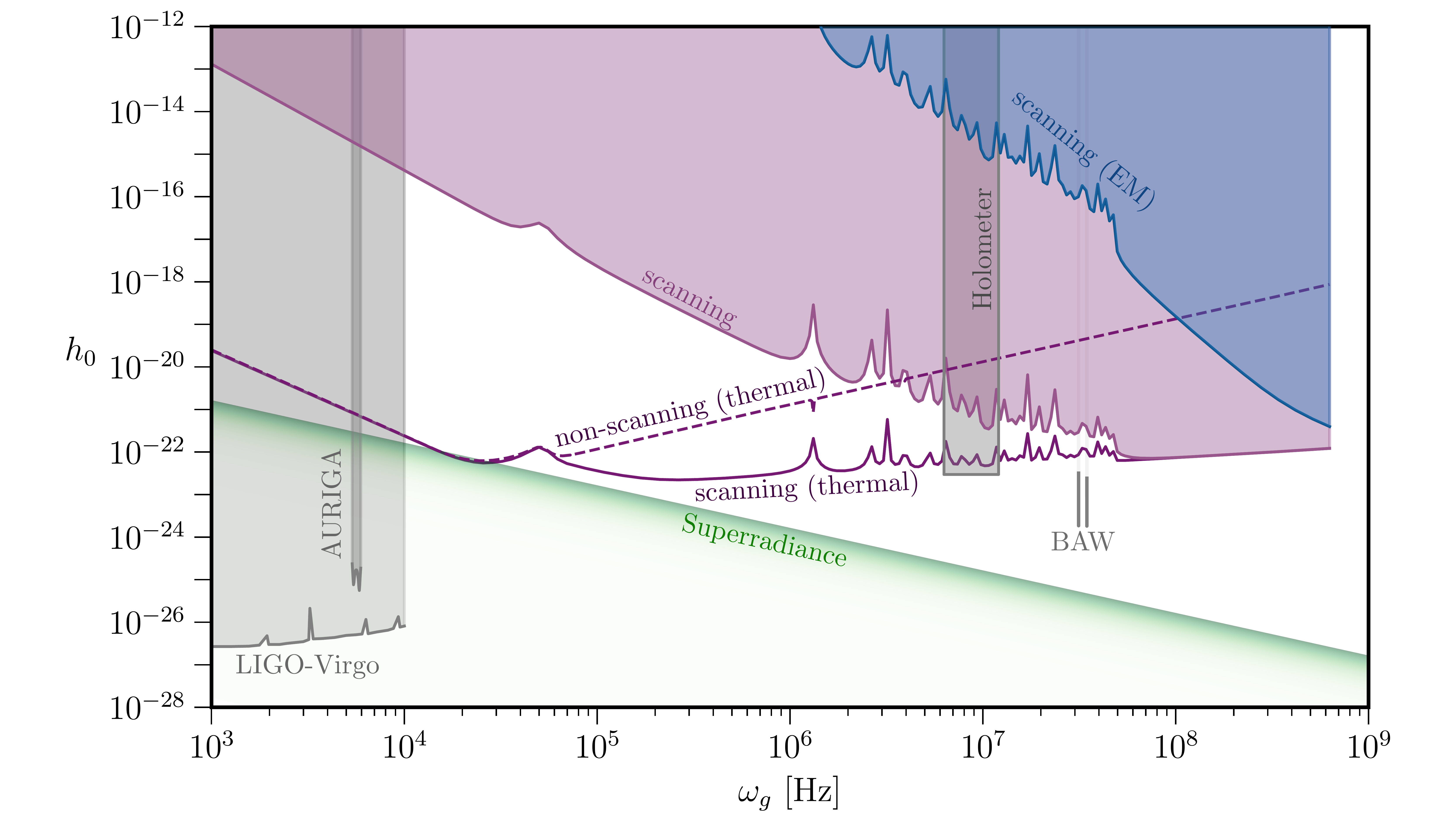}
\caption{Reach of a MAGO-like setup to coherent GWs. The mechanical (purple) and EM (blue) signals are separated for visual comparison, but they would both be present in a single experiment. The shaded purple and blue regions labeled ``scanning" and ``scanning (EM)" show the sensitivity to mechanical and EM signals, respectively, for a scanning setup in which the EM mode splitting is matched to the GW frequency, i.e., $\w_1 - \w_0 = \wg$  and assuming  vibrational noise as inferred by recent Fermilab measurements of cavity microphonics. The solid and dashed contours labeled ``scanning (thermal)" and ``non-scanning (thermal)" show the sensitivity when vibrational noise is attenuated to its irreducible thermal value, for a scanning or broadband setup, respectively. In the latter case, the EM mode splitting is fixed to the lowest-lying spin-2 mechanical resonance, i.e., $\w_1 - \w_0 = \min \w_p \sim 10 \ \text{kHz}$. In the scanning or broadband setup, the time to cover an $e$-fold in $\wg$ or the total experimental time are fixed to $1 \ \text{year}$, respectively. The degree of overcoupling to the readout is optimized for $10^5 \leq Q_\text{cpl} \leq 10^{10}$ (fixed to $Q_\text{cpl} = 10^5$) at each frequency for the scanning (non-scanning) projections. Also shown in gray are existing limits from LIGO-Virgo~\cite{LIGOScientific:2021hvc}, AURIGA~\cite{Cerdonio:1997hz,Vinante:2006uk,Branca:2016rez},
bulk acoustic wave (BAW) resonators~\cite{Goryachev:2021zzn}, and the Holometer experiment~\cite{Holometer:2016qoh}. The green shaded region corresponds to signals generated from superradiant bosonic clouds around black holes of mass $M_\star \sim M_\odot \, (10^5 \ \text{Hz}/\wg)$ at a distance of $1 \ \text{kpc}$ (see \App{sources}).}
\label{fig:Reach}
\end{figure}

In this section, we estimate the sensitivity of both a scanning and non-scanning (i.e., broadband) setup, for which the EM frequency splitting is either tuned to the GW frequency, or instead is held fixed such that a generic GW signal registers outside of the resonant bandwidth, respectively. As we show below, if external vibrations can be significantly attenuated, the existing MAGO cavity, operated in a broadband manner, could be the most sensitive instrument to GWs across a broad frequency range (see the projection labeled ``non-scanning"  in \Fig{Reach}). To compute the signal-to-noise ratio (SNR) of a MAGO-like setup, we combine all noise PSDs described in \Sec{noise} to determine the total noise PSD $S_\text{noise}(\w)$. In doing so, we include the contribution of all mechanical resonances to $S_\text{noise}^{(\text{mech})} (\w)$ (see \Fig{PSDs}).  For a GW-induced signal PSD $S_\text{sig} (\w)$, the SNR is given by integrating the ratio of signal and noise PSDs squared, 
\begin{align}
\label{eq:SNR}
\text{SNR} \simeq \Bigg[\, \frac{t_{\rm int}}{2\pi} \int_0^\infty d\w \left(\frac{S_\text{sig} (\w)}{S_\text{noise}(\w)} \right)^2 \, \Bigg]^{1/2}
~,
\end{align}
where $t_\text{int}$ is the integration time spent at a fixed frequency splitting. For a broadband search with fixed frequency splitting, $t_\text{int}$ is simply the total runtime of the experiment. Instead, in a scanning setup, $t_\text{int}$ is related to the time $t_e$ needed to cover an $e$-fold in $\wg$ by $t_\text{int} \simeq (\w_0 / \wg) \, (t_e/Q_1) \ll t_e$, assuming $\w_0 / Q_1 \ll \wg \ll \w_0$. Hence, in the latter case, $t_\text{int}$ is $\wg$-dependent for a scanning strategy employing a fixed $t_e$ (which is a sensible choice assuming a log-uniform prior for $\wg$).  In our projections, we adopt an $e$-fold time or total integration time of one year, for a scanning or broadband setup, respectively. The expression for the SNR in \Eq{eq:SNR} can be evaluated analytically in the case that a single noise source dominates. As discussed in \Sec{mechsignal}, since the GW's coupling to excited mechanical modes scales as $|\eta_\text{mech}^g| \propto 1/\wg^2$ for  $\wg \gg 10 \ \kHz$, the signal is dominated by the lowest-lying spin-2 mechanical resonance for most of the parameter space of interest. Hence, in presenting the analytic expressions below, we restrict to the case that the signal arises from a GW driving this first mechanical mode at a frequency much above its resonance, as this applies to most of the parameter space of interest and simplifies the expressions considerably. 

Let us begin by considering a mechanical-noise limited setup, which is valid for $\wg \ll \order{1} \ \text{MHz}$ (see \Fig{PSDs}). In this case, for a coherent GW the SNR in \Eq{eq:SNR} evaluates to 
\be
\label{eq:SNRmech}
\text{SNR}_\text{mech noise} \simeq 
\frac{1}{4} ~ \sqrt{\frac{\pi^3 \, t_\text{int}}{2 \, \Delta \w_\text{osc}}} ~ \, |\eta_\text{mech}^g (\w_p^\text{sig})|^2 \, \frac{|\eta_\text{mech}^\text{EM} (\w_p^\text{sig})|^2}{|\eta_\text{mech}^\text{EM} (\w_p^\text{noise})|^2} ~ \frac{\Mcav^2 \, \Vcav^{2/3}}{S_{F_p} (\wg)} ~  h_0^2 \, \wg^4
\times
\begin{cases}
1 & ,~ \wg \neq \w_p^\text{noise}
\\
1/(Q_p^{\text{noise}})^{\, 2} & ,~ \wg \simeq \w_p^\text{noise}
\, ,
\end{cases}
\ee
where we have assumed $t_\text{int} \gtrsim \Delta \w_\text{osc}^{-1}$ and in the first and second lines we have taken the GW frequency to be out or in resonance with a higher (non-lowest-lying) mechanical mode, respectively; in the latter case, the noise is resonantly enhanced, reducing the SNR by a factor of $1/(Q_p^{\text{noise}})^{\, 2}$. Note that \Eq{eq:SNRmech} distinguishes between the EM-mechanical coupling $\eta_\text{mech}^\text{EM}$ for a mechanical mode that is excited by either the GW or noise. For most of the modes considered, each of these coupling-coefficients is $\order{1}$. We also see that the mechanical-noise limited SNR scales as $\text{SNR} \propto t_\text{int}^{1/2} \, \wg^4 / S_{F_p} (\w_g)$, which is a rapidly growing function of $\wg$ in a scanning or broadband setup for most sources of vibrational noise. 

Instead, if thermal EM noise is the largest contributor, which is valid for $\wg \gtrsim \order{1} \ \text{MHz}$ in a scanning setup (see the left-panel of \Fig{PSDs}), then the mechanical signal yields an SNR that is approximately
\be
\label{eq:SNRthermal}
\text{SNR}_\text{EM noise} \simeq 
%\frac{1}{64} 
\frac{1}{16} ~ \sqrt{\frac{\pi \, t_\text{int}}{2 \, \Delta \w_\text{osc}}} ~ Q_\text{int}^2 \, |\eta_\text{mech}^g|^2 \, |\eta_\text{mech}^\text{EM}|^2 ~ \frac{P_\text{in}}{T} \, h_0^2
~.
\ee
This implies that the EM-noise limited SNR scales as $\text{SNR} \propto \sqrt{t_e / \wg}$ for a scanning search. It is also crucial to compare the dependence on the EM quality factor $Q_\text{int}$ in Eqs.~(\ref{eq:SNRmech}) and (\ref{eq:SNRthermal}). In particular, the signal in a mechanically-noise limited setup is independent of $Q_\text{int}$, as both the signal and noise are similarly enhanced by the large quality factor. However, in an EM-noise limited setup, larger $Q_\text{int}$ suppresses thermal occupation of the signal mode, such that the SNR scales as the square of the intrinsic EM quality factor for fixed input power $P_\text{in}$. The fact that EM noise is drastically reduced by $Q_\text{int} \gtrsim 10^{10}$ (or equivalently, that the EM-noise limited SNR is enhanced compared to the mechanically-noise limited one) is directly the reason why a MAGO-like setup is more sensitive to high-frequency GWs than modern Weber bars, since the latter employ electrical readout schemes with much smaller EM quality factors. We will revisit this point in more detail in the next section. 

In a broadband setup, amplifier noise dominates for $\wg \gtrsim \order{100} \ \text{kHz}$ (see the right-panel of \Fig{PSDs}). Taking $\wg \gg |\w_1 - \w_0|\, ,\, 10 \ \kHz \gg \w_0 / Q_1$, the approximate SNR of an amplifier-noise limited broadband experiment is 
\be
\label{eq:SNRamp}
\text{SNR}_\text{amp noise}^{(\text{broadband})} \simeq %\frac{1}{64} 
\frac{1}{16} ~ \sqrt{\frac{\pi \, t_\text{int}}{2 \, \Delta \w_\text{osc}}} ~ \frac{Q_\text{int}}{Q_\text{cpl}} \, |\eta_\text{mech}^g|^2 \, |\eta_\text{mech}^\text{EM}|^2 ~ \frac{P_\text{in}}{T} \, h_0^2 \, \frac{\w_0}{\wg^2}
~.
\ee
We see from \Eq{eq:SNRamp} that the SNR decreases rapidly with increasing $\wg$, since here the GW excites the signal mode off-resonance (i.e., $\wg \gg |\w_1 - \w_0|$), whereas the amplifier noise is independent of the EM resonance. Unlike the previous two examples, in this case overcoupling the signal mode to the readout $Q_\text{cpl} \ll Q_\text{int}$ parametrically increases the SNR, since amplifier noise is assumed to remain independent of $Q_\text{cpl}$. In our projections for a scanning experiment, we optimize the SNR by marginalizing over $Q_\text{cpl}$ for each value of $\wg$, while demanding $Q_\text{cpl} \geq 10^5$~\cite{Berlin:2020vrk}. Instead, for a broadband experiment we fix $Q_\text{cpl} = 10^5$. Comparing Eqs.~(\ref{eq:SNRmech}), (\ref{eq:SNRthermal}), and (\ref{eq:SNRamp}), we see that the impact of various experimental parameters (such as the pump mode power, quality factors, and vibration attenuation) depends sensitively on the nature of the dominant noise source and hence the particular frequency range considered. For instance, if mechanical vibrations dominate the noise budget, as in \Eq{eq:SNRmech}, the SNR is independent of the EM quality factor and pump mode power; varying these parameters affects both the signal and noise in the same way since they enter through the same physical process. On the other hand, these very same parameters are critically important for enhancing the sensitivity when limited by EM thermal noise, as in \Eq{eq:SNRthermal}. 

The above expressions provide analytic handles for understanding the sensitivity to coherent GWs in various frequency/noise regimes. In our actual estimates, we employ a numerical evaluation of \Eq{eq:SNR}, incorporating all noise sources previously discussed. Our results for the sensitivity (corresponding to $\text{SNR} \geq 1$) to coherent GW sources (i.e., GWs that are more coherent than the external oscillator used to drive the pump mode) are shown in \Fig{Reach} for a various experimental setups. The solid purple lines correspond to a scanning experiment where the splitting between the EM modes is tuned to the GW frequency at every point (note that the  mechanical modes are held fixed). For the line labeled ``scanning," we adopt the vibrational noise of \Eq{eq:fermilabforce}, estimated from recent cavity microphonics measurements at Fermilab. Instead, we assume that such noise has been attenuated down to the irreducible thermal value of \Eq{eq:thermalforce} for the projection labeled ``scanning (thermal),"
the feasibility of which is discussed above in \Sec{noise}. The dashed purple line labeled ``non-scanning (thermal)" demonstrates the sensitivity of a broadband setup employing an EM mode splitting fixed to the frequency of the lowest-lying spin-2 mechanical resonance, i.e., $\w_1 - \w_0 \simeq \min{\w_p} \sim 10 \ \kHz$, and, once again, assuming vibrational noise attenuated to its irreducible thermal value.\footnote{We have chosen this value for the EM mode splitting in a broadband setup because it optimizes the reach across a wide a range of GW frequencies and roughly matches the central frequency splitting of the existing MAGO prototype.} 
Along the blue solid line labeled ``scanning (EM)", we also show the sensitivity of a scanning setup limited by the vibrational noise given in \Eq{eq:fermilabforce} and targeting the direct EM signal described in \Sec{EMsignal}. As discussed above, this signal is parametrically suppressed compared to the mechanical one for $ \wg \ll \w_0$, leading to a drastically reduced sensitivity at small frequencies. Also shown in \Fig{Reach} are existing limits (solid gray) on coherent high-frequency GWs. These include searches performed by LIGO and VIRGO\footnote{Note that the reach of LIGO-Virgo cannot be extrapolated beyond $\sim 10 \ \text{kHz}$. Besides the fact that current data is sampled at $\sim 16 \ \text{kHz}$, there is also a lack of feasible calibration to understand and control the changed optical response at high frequencies. We thank M. Seglar and O. Piccinni for discussions on this point.}~\cite{LIGOScientific:2021hvc}, the Weber bar experiment AURIGA~\cite{Cerdonio:1997hz,Vinante:2006uk,Branca:2016rez}, the Holometer interferometer~\cite{Holometer:2016qoh}, and a bulk acoustic wave (BAW) resonant mass antenna~\cite{Goryachev:2021zzn}. The green shaded region corresponds to the predicted signal strength of coherent GWs generated from superradiant bosonic clouds around black holes of mass $M_\star \sim M_\odot \, (10^5 \ \text{Hz}/\wg)$ at a distance of $1 \ \text{kpc}$~\cite{Arvanitaki:2009fg,Arvanitaki:2010sy}. We refer the interested reader to \App{sources} for further discussion of such signals.

\section{Comparisons to Other Experiments}
\label{sec:comparison}

\begin{figure}
    \centering
    \includegraphics[width = 0.8\textwidth]{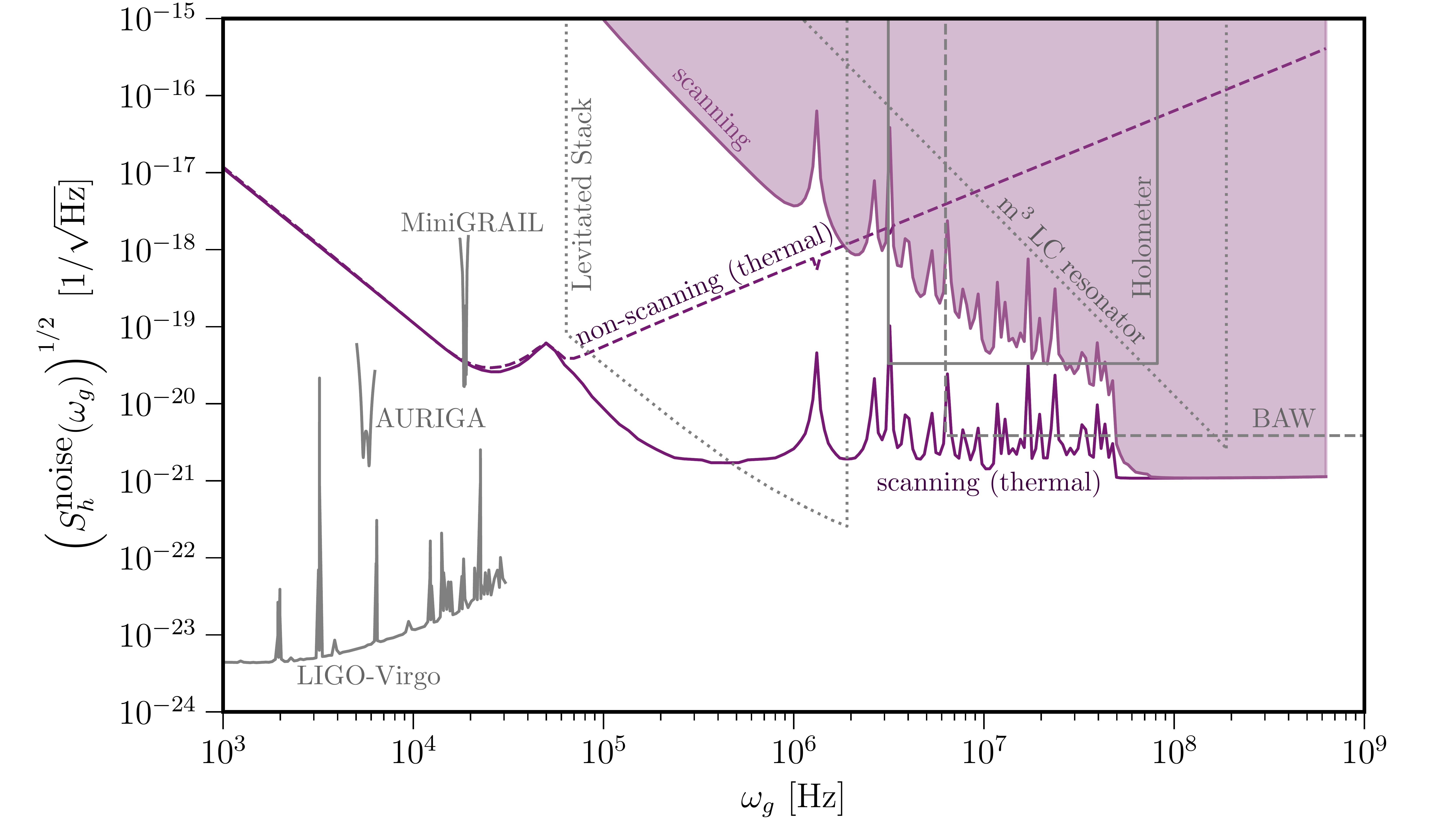}
    \caption{The strain-equivalent noise $S_h^\text{noise}$, as discussed in \Sec{comparison}. The purple regions and contours correspond to such strain noise for mechanical signals in a MAGO-like setup in the same configurations as discussed in \Fig{Reach}. The solid gray lines correspond to  measured strain noise of existing experiments, such as LIGO-Virgo~\cite{LIGOScientific:2021hvc}, AURIGA~\cite{Cerdonio:1997hz,Vinante:2006uk}, MiniGRAIL~\cite{Gottardi:2007zn}, and the Holometer experiment~\cite{Holometer:2016qoh}. In dashed or dotted gray are projections of strain noise in bulk acoustic wave (BAW) resonators~\cite{Goryachev:2021zzn} and other proposals including meter-sized levitated dielectric stacks~\cite{Aggarwal:2020umq} and LC circuits (assuming parameters comparable to the upcoming DMRadio-m$^3$ axion dark matter experiment~\cite{DMRadio:2022pkf} and a specialized readout architecture~\cite{Domcke:2022rgu}).}
    \label{fig:SHReach}
\end{figure}

In \Fig{Reach}, we estimated the reach to coherent GWs with amplitude $h_0$. However, to compare to other experimental setups, it is often more useful to phrase the sensitivity in terms of the ``effective noise strain" PSD $S_h^\text{noise}$, since it is independent of choices regarding scan strategies and observation time, and it can be used to determine the reach by noting that the ratio of signal and noise PSDs is $S_\text{sig} / S_\text{noise} = S_h / S_h^\text{noise}$, where $S_h$ is the PSD of the GW strain $h^\text{TT}$ in the TT frame. To determine $S_h^\text{noise}$, we define the ``transfer function" $\mathcal{T}$ as the ratio between the signal power and GW PSDs, $S_\text{sig} (\w_0 + \wg) = \mathcal{T} \, S_h (\wg)$, for a \emph{non-coherent} GW source (i.e., one whose coherence time is much shorter than the cavity ringup time). Since \Eq{eq:sigPSD} was derived assuming a monochromatic GW source, $S_\text{sig}$ must be rederived assuming a spectrally broad signal. This is done in detail in \App{mechPSDs}, which shows that the transfer function of a MAGO-like experiment is given by 
\be
\mathcal{T} \simeq %\frac{P_\text{in}}{16} 
\frac{P_\text{in}}{4} \, \frac{Q_\text{int}}{Q_\text{cpl}} ~\frac{\w_0^4 ~  |\eta_\text{mech}^\text{EM} (\w_p^\text{sig})|^2}{\big[(\w_0 + \wg)^2 - \w_1^2\big]^2 + \big[(\w_0 + \wg) \, \w_1 / Q_1 \big]^2} ~ \frac{\wg^4 ~ |\eta_\text{mech}^g (\w_p^\text{sig})|^2}{(\wg^2 - \w_p^{\text{sig} \, 2} )^2 + (\wg \, \w_p^\text{sig} / Q_p^\text{sig} )^2}
~,
\label{eq:transferFunction}
\ee
where $\w_p^\text{sig}$ and $Q_p^\text{sig}$ are fixed to the frequency and quality factor of the lowest-lying spin-2 mechanical resonance. The expression above can be used to convert the noise power PSDs of \Sec{noise} to effective noise strain PSDs, $S_h^\text{noise} (\wg) = S_\text{noise}(\w_0 + \wg) \, / \, \mathcal{T}$. This is done in \Fig{SHReach}, which shows the projections for existing and proposed experimental setups phrased in terms of $(S_h^\text{noise})^{1/2}$. 

Similar to Eqs.~(\ref{eq:SNRmech})$-$(\ref{eq:SNRamp}), we give analytic results for $S_h^\text{noise}$ in the case that a single noise source dominates and that the signal arises from a GW driving the first mechanical mode at a frequency much above its resonance, $\wg \gg \w_p^\text{sig} \sim 10 \ \text{kHz}$. Analogous to \Eq{eq:SNRmech}, for a mechanical-noise limited setup (e.g., for $\wg \ll 1 \ \text{MHz}$), 
\be
S_h^\text{mech noise} (\wg) \simeq \frac{4 / \wg^4}{|\eta_\text{mech}^g (\w_p^\text{sig})|^2}~ \frac{|\eta_\text{mech}^\text{EM} (\w_p^\text{noise})|^2}{|\eta_\text{mech}^\text{EM} (\w_p^\text{sig})|^2} ~ \frac{S_{F_p} (\wg)}{\Mcav^2 \, \Vcav^{2/3}} 
\times
\begin{cases}
1 & ,~ \wg \neq \w_p
\\
Q_p^2 & ,~ \wg \simeq \w_p
~,
\end{cases}
\label{eq:shmechnoise}
\ee
where in the first and second lines we have taken the GW frequency to be out or in resonance with a higher (non-lowest-lying) mechanical mode, respectively. Instead, for a thermal EM-noise limited setup analogous to \Eq{eq:SNRthermal} (e.g., for $\wg \gtrsim 1 \ \text{MHz}$ in a scanning experiment), the effective noise strain is
\be
\label{eq:ShEMnoise}
S_h^\text{EM noise}(\wg) \simeq \frac{%64 
16 \pi \, T}{|\eta^\text{EM}_\text{mech}|^2 \, |\eta^g_\text{mech}|^2 \, Q_\text{int}^2 \,  P_\text{in}}
~.
\ee
Finally, when amplifier noise dominates (e.g., for $\wg \gtrsim 100 \ \text{kHz}$ in a broadband setup), analogous to \Eq{eq:SNRamp} we have
\be
S_h^\text{amp noise} (\wg) \simeq \frac{%64
16 \pi \, Q_\text{cpl} \, \wg^2}{P_\text{in} \, Q_\text{int} \, \w_0 \, |\eta^\text{EM}_\text{mech}|^2 \, |\eta^g_\text{mech}|^2} 
~,
\label{eq:shAmpNoise}
\ee
where we took $\wg \gg |\w_1 - \w_0|\, ,\, 10 \ \kHz \gg \w_0 / Q_1$. The frequency-scaling of the projections in \Fig{SHReach} follow straightforwardly from these analytic expressions. In our actual calculations, we incorporate all noise sources previously discussed. In \Fig{SHReach}, we also show the same limits (solid gray) as shown previously in \Fig{Reach}, but now translated to $S_h^\text{noise}$. Also shown are projections (dashed or dotted gray) of other experiments sensitive to high-frequency GWs. These include the projected sensitivity of aLIGO~\cite{ligo}, Weber bar antennas~\cite{Cerdonio:1997hz, Gottardi:2007zn}, the Holometer experiment~\cite{Holometer:2016qoh}, levitated dielectric stacks~\cite{Aggarwal:2020umq}, a modified version of DMRadio~\cite{Domcke:2022rgu} (see \Sec{LCcompare}), and bulk acoustic wave (BAW) resonant mass antennae~\cite{Goryachev:2021zzn}. In comparison, we see that a tunable MAGO-like instrument solely limited by irreducible vibrational noise would be the most powerful tabletop-sized device across a large frequency range. 

\subsection{Weber Bars}
\label{sec:Webercompare}

We conclude this section with a more detailed comparison to signals at Weber bar and LC circuit experiments. As mentioned above, the setup discussed throughout this work is most similar in spirit to a Weber bar experiment. Indeed, both instruments act as mechanical to EM transducers, as modern-day Weber bars have typically employed low-noise LC circuits to capacitively readout small mechanical displacements. Although the hollow cavity discussed for MAGO is much less massive than a typical ton-scale resonant antenna, the more important difference arises from the fact that the EM thermal noise of LC circuits (which dominates outside of the bandwidth of the mechanical resonance) is parametrically larger than that of a superconducting cavity. 

To see this in detail, let us consider the equations of motion governing the signal and readout of a Weber bar. These are given by the coupled set of equations describing the displacement $u_b$ of the bar, the displacement $u_t$ of the smaller mechanical transducer, and the charge $q$ of the circuit capacitor. In particular, the bar and transducer are modeled as a mechanically-coupled double oscillator with masses $M_b \gg M_t$ and uncoupled resonant frequencies $\w_b$ and $\w_t$, respectively. This mechanical system capacitively couples to the LC circuit, where the position of the mechanical transducer modulates a free plate of the circuit capacitor to which a bias static electric field of $E_C \sim 10 \ \text{MV} / \text{m}$ is applied. Hence, mechanical displacements of the capacitor plate drive voltage through the circuit, and in return the electric force on the capacitor backreacts with a mechanical force on the transducer and bar. The coupled equations of motion are given by ~\cite{Tricarico:1993by,Vinante:2006uk,Maggiore}
\begin{align}
& M_b \, \ddot{u}_b + \frac{M_b \, \w_b}{Q_b} \, \dot{u}_b + M_b \, \w_b^2 \, u_b + \frac{M_t \, \w_t}{Q_t} \, (\dot{u}_b - \dot{u}_t) + M_t \, \w_t^2 \, (u_b - u_t) + E_C \, q = F_b
\nl
& M_t \, \ddot{u}_t + \frac{M_t \, \w_t}{Q_t} \, (\dot{u}_t - \dot{u}_b) + M_t \, \w_t^2 \, (u_t - u_b) - E_C \, q = F_t
\nl
& L \, \ddot{q} + \frac{L \, \w_\text{LC}}{Q_\text{LC}} \, \dot{q} + \frac{q}{C} + E_C \, (u_t - u_b) = V_\text{LC}
~,
\end{align}
where $F_b$ is the force coupled to the bar (including both the GW and noise contributions), $F_t$ is the force coupled to the mechanical transducer (approximately just from noise), $V_\text{LC}$ is the voltage associated with thermal noise fluctuations, $\w_{b, t, \text{LC}}$ and $Q_{b, t, \text{LC}} \sim 10^6$ are the resonant frequency and the quality factor of the bar, transducer, and circuit, and $L$ is the inductance of the circuit. 

To proceed, we Fourier transform the above equations in order to solve for the  power delivered to the circuit by the GW signal or noise, taking the transducer to be very light $M_t \ll M_b$. It is useful to compare the irreducible noise from thermal occupation of EM modes of the circuit or mechanical modes of the bar/transducer,
\be
\label{eq:WeberEMtoMech}
\frac{S_\text{noise}^{(\text{EM})} (\w)}{S_\text{noise}^{(\text{mech})} (\w)} \Bigg|_\text{Weber} \hspace{-0.2cm} \sim ~ \frac{Q_b \, M_t}{Q_\text{LC} \, E_C^2 \, C} ~ \frac{(\w^2 - \w_b^2)^2 + (\w \, \w_b/ Q_b)^2}{\w \, \w_\text{LC}}
~,
\ee
where $C \sim 10 \ \text{nF} \sim 1 \ \text{km}$ is the capacitance of the circuit, and we have taken $\w_b \sim \w_t$ and $Q_b \sim Q_t$. From \Sec{noise}, the analogous ratio of noise PSDs derived for the MAGO-like setup is
\be
\label{eq:MAGOEMtoMech}
\frac{S_\text{noise}^{(\text{EM})} (\w_1)}{S_\text{noise}^{(\text{mech})} (\w_1)} \Bigg|_\text{MAGO} \hspace{-0.2cm} \sim ~ \frac{Q_p \, M_\text{cav}}{Q_\text{int} \, E_0^2 \, \Vcav^{1/3}} ~ \frac{(\Delta \w_\text{EM}^2 - \w_p^2)^2 + (\Delta \w_\text{EM} \, \w_p/ Q_p)^2}{\w_p \, \w_0}
~,
\ee
where $\Delta \w_\text{EM} \equiv \w_1 - \w_0$ is the EM mode splitting.  Eqs.~(\ref{eq:WeberEMtoMech}) and (\ref{eq:MAGOEMtoMech}) illustrate the relative size of irreducible EM or mechanical noise in a Weber bar or MAGO-like setup, respectively. In comparing the two results, the largest difference arises from the fact that the typical EM quality factor of superconducting cavities $Q_\text{int} \gtrsim 10^{10}$ is parametrically larger than what has been achieved in lumped element circuits  $Q_\text{LC} \sim 10^6$. In particular, the first factor in \Eq{eq:MAGOEMtoMech} is smaller than the first factor of  \Eq{eq:WeberEMtoMech} by a factor of $\sim 10 - 10^3$ for $Q_\text{int} \sim 10^{10} - 10^{12}$ and fixing the other experimental parameters to those specified in \Sec{sensitivity}. As a result, EM noise is greatly suppressed for superconducting cavities, enabling mechanical-noise limited sensitivity to GWs well outside the mechanical resonance and drastically enhancing the reach to GW frequencies $\wg \gg 10 \ \kHz$ compared to traditional Weber bar experiments. Note, however, the smaller mass and volume of superconducting cavities implies a reduced sensitivity to GWs on resonance with a low-lying mechanical mode, in which case the dominant source of noise arises from mechanical vibrations.  

\subsection{LC Resonators}
\label{sec:LCcompare}

Ref.~\cite{Domcke:2022rgu} demonstrated that LC circuit resonators, similar to the ones being developed to search for axion dark matter, can also search for high-frequency GWs. In particular, Ref.~\cite{Domcke:2022rgu} calculated the sensitivity to the GW amplitude $h_0$ for a particular set of assumptions regarding scanning strategy, integration time, and GW coherence time. We cannot compare directly our sensitivity to Ref.~\cite{Domcke:2022rgu} since they assumed GW waveforms of a particular type and their adopted scan strategy is dictated by the proposed DMRadio-GUT~\cite{DMRadio:2022jfv} search for QCD axion dark matter, which results in longer integration times for lower frequency signals. To avoid misrepresenting the sensitivity of such an experiment, we estimate here the irreducible capability of a future cubic meter LC resonator, adopting a scan strategy employing a fixed $e$-fold time $t_e$. As described in \Sec{EMsignal}, an LC circuit detector is sensitive to the EM signal generated by the effective current $j_\text{eff}$ of \Eq{eq:jeff}, which is parametrically of the form $j_\text{eff} \sim h_0 \, B_{\rm LC} \, \wg^2 \, 
\VLC^{1/3}$, where $\VLC$ is the volume of the circuit and $B_{\rm LC}$ is the  static magnetic field applied to the setup. The effective current sources an oscillating magnetic field $B_h \sim j_\text{eff} \, \VLC^{1/3} \, e^{i \wg t}$, generating an oscillating electromotive force $\mathcal{E}_h \sim \eta_\text{LC}^g \, \wg \, B_h \, N_\text{coil} \, \VLC^{2/3}$ through a large pickup inductor employing $N_\text{coil}$ turns of superconducting wire, where $\eta_\text{LC}^g \leq 1$ is a dimensionless coefficient that parametrizes the geometric coupling of the GW to the pickup inductor ($\eta_\text{LC}^g \sim \order{1}$ is possible with a specialized ``figure-8" pickup loop~\cite{Domcke:2022rgu}).  The resulting signal power driven into the circuit is described by the PSD
\be
S_\text{sig}^{(\text{LC})} (\w) =  
\frac{(\w \, \wLC / Q_\text{LC})^2}{(\w^2 - \wLC^2)^2 + (\w \, \wLC / Q_\text{LC})^2}
~ \frac{Q_\text{LC}}{\wLC \, L} ~ S_{\mathcal{E}_h}(\w)
= \mathcal{T}_\text{LC} ~ S_{h}(\w)
~,
\ee
where $\wLC$, $Q_\text{LC}$, and $L$ are the resonant frequency, quality factor, and inductance of the circuit, respectively, and we have defined the transfer function
\be
\mathcal{T}_\text{LC} \sim \frac{(\w \, \wLC / Q_\text{LC})^2}{(\w^2 - \wLC^2)^2 + (\w \, \wLC / Q_\text{LC})^2}
~ \frac{Q_\text{LC}}{\wLC} \, |\eta_\text{LC}^g|^2 \, \wg^6 \, \VLC^{7/3} \, B_{\rm LC}^2
~.
\ee
Analogous to \Eq{eq:EMthermalPSD}, the PSD for thermal noise in the circuit is given by
\be
S_\text{noise}^{(\text{LC})} (\w) =  \frac{4 \pi T \, (\w \, \wLC / Q_\text{LC})^2}{(\w^2 - \wLC^2)^2 + (\w \, \wLC / Q_\text{LC})^2}
~.
\ee
Dividing the above expression by the transfer function $\mathcal{T}_\text{LC}$ gives the effective noise strain of a thermal-noise limited LC circuit detector, 
\be
S_h^{(\text{LC noise})} (\wg) \sim
\frac{4 \pi T}{|\eta_\text{LC}^g|^2 \, Q_\text{LC} \, \wg^5 \, \VLC^{7/3} \, B_{\rm LC}^2}
~,
\ee
where we have taken the circuit to be tuned to the GW frequency, $\wLC \simeq \wg$. Comparing the above expression to the corresponding sensitivity $S_h^\text{EM noise}$ of a thermal EM-noise limited MAGO-like setup in \Eq{eq:ShEMnoise},
\be
\frac{S_h^{(\text{LC noise})} (\wg)}{S_h^\text{EM noise}(\wg)} \propto \bigg( \frac{B_0}{B_\text{LC}} \bigg)^2 \, \bigg( \frac{Q_\text{int}}{Q_\text{LC}}\bigg) \, \bigg( \frac{\w_0}{\wg} \bigg)^5
~,
\ee
we see that the smaller magnetic field in superconducting cavities $B_0 / B_\text{LC} \sim 10^{-2}$ is easily compensated by the larger quality factor $Q_\text{int} / Q_\text{LC} \sim 10^4$ and the optimized frequency scaling in the quasistatic limit $\wg \ll \w_0 \sim \Vcav^{-1/3}$. This is evident in \Fig{SHReach}, which shows that when comparing these two setups, the irreducible noise in a MAGO-like instrument is parametrically smaller for $\wg \ll 1 \ \text{GHz}$. This discussion regarding the effective noise strain directly implies that a thermal-noise limited superconducting heterodyne search would have greater sensitivity to spectrally broad GWs. We note, however, that this same conclusion also applies to coherent GW signals. In particular, using the formalism presented previously in \Sec{mechsignal}, we find that a MAGO-like setup would have enhanced sensitivity to coherent GWs with frequency $\wg \lesssim 100 \ \text{MHz}$.

\section{Outlook}
\label{sec:outlook}

In this work, we have revisited heterodyne signals of high-frequency GWs in superconducting cavities, which operate analogously to modern-day Weber bars but with parametrically reduced EM noise in the readout. This was pursued by the MAGO collaboration in the early 2000s, which led to a prototype cavity and readout system. Our analysis extends that of previous studies performed by the MAGO collaboration in a number of ways, which motivates revisiting the experiment. For instance, we have provided a more detailed analysis of the most relevant noise sources as well as the EM and mechanical signals, which demonstrates that a broadband run of this same prototype is capable of exploring orders of magnitude of new parameter space. More generally, our study provides guidance to a future dedicated effort. In particular, if further designs could significantly mitigate external sources of vibrations and also allow for tunable EM mode separations across a wider frequency range, an EM-resonant experiment would have nearly frequency-independent sensitivity to strains of $h_0 \sim 10^{-22}$ for GW frequencies of $10 \ \text{kHz} - 1 \ \text{GHz}$.

This effort is quite timely both from an experimental and theoretical perspective. Experimentally, there is a vast ongoing effort on both sides of the Atlantic (e.g., at Fermilab's SQMS center, CERN's Quantum Technology Initiative and RF department, and DESY's accelerator division) aiming to improve all aspects of design, construction, and operation of superconducting cavities. In regards to theory, the first detection of GWs by the LIGO collaboration has ushered in a series of predictions for new sources, including at the higher frequencies that we consider in this work~\cite{Ghiglieri:2020mhm,Ringwald:2020ist,Ghiglieri:2022rfp,Casalderrey-Solana:2022rrn, Franciolini:2022htd,Franciolini:2022ewd}. Our improved analysis shows that the potential sensitivity of MAGO to such signals is greater than that envisaged at the time of its initial proposal, making the 
the physics case for reviving this experiment (and pursuing related technology~\cite{Lorenzo_2014,Singh:2016xwa,Lorenzo:2016isc,Manley:2019vxy,Vadakkumbatt:2021fnw}) stronger than ever before.

\acknowledgements

We thank Alex Dima, Valerie Domcke, Camilo Garcia-Cely, Sam Posen, Nicholas Rodd, Vyacheslav Yakovlev, and Kevin Zhou for helpful discussions. This  material  is  based  upon  work  supported  by  the U.S.\ Department of Energy,  Office of Science,  National Quantum  Information  Science  Research  Centers,   Superconducting  Quantum  Materials  and  Systems  Center (SQMS) under contract number DE-AC02-07CH11359. 
Fermilab is operated by the Fermi Research Alliance, LLC under Contract DE-AC02-07CH11359 with the U.S.\ Department of Energy. The work of SARE was supported by SNF Ambizione grant PZ00P2\_193322, \textit{New frontiers from sub-eV to super-TeV}. SARE and JSE thank the MITP for hospitality while part of this work was conducted. The work of YK, JSE, and MW is supported in part by DOE grant DE-SC0015655. DB is supported by a `Ayuda Beatriz Galindo Senior' from the Spanish `Ministerio de Universidades', grant BG20/00228. 
The research leading to these results has received funding from the Spanish Ministry of Science and Innovation (PID2020-115845GB-I00/AEI/10.13039/501100011033).
IFAE is partially funded by the CERCA program of the Generalitat de Catalunya. This material is based upon work supported by the National Science Foundation Graduate Research Fellowship Program under Grant No. DGE 21-46756.

\renewcommand\thefigure{\thesection.\arabic{figure}}   
\appendix
\addappheadtotoc
\setcounter{figure}{0} 
\section{Mechanical Modes}
\label{app:mechanical-modes}

Results for the mechanical resonant modes of hollow spheres have been obtained in Ref.~\cite{Coccia:1997gy}. We summarize these results here, and in so doing, correct a couple of typographical errors. The equation of motion for displacements in a material is given by~\cite{Landau:1986aog} 
\begin{align}
\rho \, \pd_t^2 \mathbf{U} = (\lambda+\mu) \nabla(\nabla\cdot \mathbf{U}) + \mu \nabla^2 \mathbf{U} + \mathbf{f}_{\rm ext} \ ,
\label{eq:mechEOM}
\end{align}
where the material properties are the density $\rho$ and the Lam\'e coefficients $\lambda$ and $\mu$. We have also included the possibility of an external force density $\mathbf{f}_{\rm ext}$ coupling to the displacement $\mathbf{U}$.  Setting this external force to zero, and assuming solutions of the form $\mathbf{U}(\mathbf{x},t) = \mathbf{U}(\mathbf{x}) \, e^{i \w t}$, we can then define the wavenumber $k \equiv \sqrt{\omega^2 \rho/\mu}$ such that the equation of motion reads
\begin{align}
\nabla^2\mathbf{U} + (1+\lambda/\mu) \nabla(\nabla \cdot \mathbf{U}) = - k^2 \mathbf{U} \ .
\end{align}
We now follow Ref.~\cite{Maggiore} and decompose $\mathbf{U} = \mathbf{U}_L + \mathbf{U}_T$, where the two vectors correspond to longitudinal (L) and transverse (T) motion. These two vectors obey $\nabla \times \mathbf{U}_L = 0$ and $\nabla \cdot \mathbf{U}_T = 0$, respectively. This decomposition allows us to write the equation of motion above as two separate equations, one for each of the L and T modes,
\begin{align}
\nabla^2 \mathbf{U}_L &= -\frac{k^2 \mu}{\lambda+2\mu}\mathbf{U}_L \ ,\\
\nabla^2 \mathbf{U}_T &= - k^2 \mathbf{U}_T \ ,
\end{align}
from which we further define $q^2 \equiv k^2 \mu/(\lambda + 2\mu)$.\footnote{Note that Ref.~\cite{Coccia:1997gy} has a typographical error in their definition of $q$ below their Eq.~(2.6).} 

Since the longitudinal component has no curl, we can write $\mathbf{U}_L = \nabla \phi_L$. The transverse component has vanishing divergence, and can be decomposed into two independent vectors $\mathbf{U}_{T_1} = i \nabla \times \mathbf{L} \phi_T,~\mathbf{U}_{T_2} = i\mathbf{L} \phi_T$, with the angular momentum operator defined as $\mathbf{L} \equiv - i \mathbf{x} \times \nabla$. The $T_1$ transverse component has radial and angular dependence, while the $T_2$ component is purely angular. The scalar potentials $\phi_L,~\phi_T$ then satisfy the Helmholtz equation,
\begin{align}
\nabla^2 \phi_L &= -q^2\phi_L \ ,\\
\nabla^2 \phi_T &= -k^2\phi_T \ .
\end{align}
The solutions to these equations in a spherical geometry are of the form
\begin{align}
\phi_L &= (c_L^i y_l( q_{nl} r) +c_L^o j_l( q_{nl} r))Y_{l m}(\theta,\p) \ , \\
\phi_{T_1} &= (c_{T_1}^i y_l( k_{nl} r) +c_{T_1}^o j_l( k_{nl} r))Y_{l m}(\theta,\p) \ , \\
\phi_{T_2} &= (c_{T_2}^i y_l( k_{nl} r) +c_{T_2}^o j_l( k_{nl} r))Y_{l m}(\theta,\p) \ .
\label{eq:LTpotentials}
\end{align}
The functions $j_l(z),~y_l(z)$ are spherical Bessel functions of the first and second kind respectively, while the $Y_{lm}(\theta,\p)$ are spherical harmonics. The constants $c^{i,o}_{L,T_1,T_2}$ are determined by the inner ($i$) and outer ($o$) boundary conditions as discussed below. This allows us to write the mechanical modes as
\begin{align}
\mathbf{U}_{lmn} = \nabla \phi_L+ i\, \nabla \times \mathbf{L} \phi_{T_1} + i\,  \mathbf{L} \phi_{T_2} \ .
\label{eq:fullMechModeEq}
\end{align}
As discussed in \App{GWmechOverlap}, only spheroidal modes  
($c_L^{i,o},~c_{T_1}^{i,o} \neq 0,~c_{T_2}^{i,o} =0 $) couple to GWs. Therefore, in what follows we will eventually set $c_{T_2}^{i,o} = 0$. 

\subsection{Boundary conditions and solutions}

To solve for the wavenumber $k$ of a given mechanical mode, we need to find the standing wave solutions of \Eq{eq:fullMechModeEq} that satisfy the boundary conditions at the inner ($a$) and outer ($R$) radii of the spherical shell. These can be derived from the conservation of the stress-energy tensor $\pd_\mu T^{\mu \nu} = 0$. The stress tensor $\sigma_{ij}$ is the spatial part of $T^{\mu\nu}$ and is defined as
\begin{align}
\sigma_{ij} = 2 \mu \varepsilon_{ij} + \lambda \delta_{ij} \sum_k \varepsilon_{kk}
~~,~~
\varepsilon_{ij} = \frac{1}{2} \left(\frac{\pd U_i}{\pd x_j} + \frac{\pd U_j}{\pd x_i} \right) \ .
\label{eq:StressStrainTensors}
\end{align}
Conservation of the stress-energy tensor implies $\pd_j \sigma_{ij} = 0$ in equilibrium. Therefore, the relevant boundary conditions for the mechanical modes are
\begin{align}
\sigma_{ij}\Big\vert_{r = R,~ r= a} n_j = 0 \ ,
\end{align}
where $n_j$ is the $j^\text{th}$ component of the unit vector $\hat{n}$ normal to the surface of the material. The boundary condition can be written as
\begin{align}
 \lambda (\nabla \cdot \mathbf{U}) \hat{\mathbf{r}} + 2 \mu (\hat{\mathbf{r}} \cdot \nabla)\mathbf{U} + \mu \hat{\mathbf{r}} \times (\nabla \times \mathbf{U}) = 0 \ .
 \label{eq:MechBC}
\end{align}
Recalling the form of the normal mechanical modes in \Eq{eq:fullMechModeEq}, we can write \Eq{eq:MechBC} as a $6\times6$ matrix multiplying the 6 unknown coefficients, $c_L^o,\, c_{T_1}^o,\,c_L^i,\,c_{T_1}^i,\,c_{T_2}^o,\,c_{T_2}^i$. The rows correspond to taking the $j=1,\,2,\,3$ components of \Eq{eq:MechBC} evaluated at $r=a,\,R$. The $6\times6$ matrix can be brought into block-diagonal form, with a $4\times 4$ matrix $\mathbf{M}$ multiplying the vector $\mathbf{C} = (c_L^o,~ c_{T_1}^o,~c_L^i,~c_{T_1}^i)^T$, and a $2\times 2$ matrix $\mathbf{N}$ multiplying the vector $\mathbf{D} = (c_{T_2}^o,~c_{T_2}^i)^T$. Since the vector $\mathbf{D}$ corresponds to the coefficients of the toroidal modes, and these do not couple to GWs, we will consider only the $4\times4$ sub-matrix $\mathbf{M}$ in what follows. The elements of $\mathbf{M}$ are given by
\begin{align}
\nonumber M_{11} &= q^2 \left(2 \mu  j_l''(a q)-\lambda  j_l(a q)\right) & M_{31} & = \frac{2 \mu  \left(a q j_l'(a q)-j_l(a q)\right)}{a^2}\\
\nonumber M_{12} &= \frac{2 l (l+1) \mu  \left(j_l(a k)-a k j_l'(a k)\right)}{a^2} & M_{32} & = -k^2 \mu  j_l''(a k)-\frac{\left(l(l+1)-2\right) \mu  j_l(a k)}{a^2}\\
M_{13} &= M_{11}|_{j_l\to y_l} & M_{33} &= M_{31}|_{j_l\to y_l} \label{eq:MatrixElements} \\
\nonumber M_{14} &= M_{12}|_{j_l\to y_l} & M_{34} &= M_{32}|_{j_l\to y_l} \\
\nonumber M_{2i} &= M_{1i}|_{a\to R} & M_{4i} &= M_{3i}|_{a\to R}
~.
\end{align}
Comparing with Ref.~\cite{Coccia:1997gy}, we find that our matrix is identical to theirs up to an arbitrary rescaling, modulo a typo in their Eq.~(2.11), where their function $\beta_2(z)$ should be defined as $\beta_2(z) = j_l''(z)$, instead of $\beta_2(z) = j_l'(z)$ as stated in that work. The wavenumbers $k_{lmn}$ of the normal modes $\mathbf{U}_{lmn}$ of the cavity are then found by solving $\det \mathbf{M}=0$. The corresponding eigenfrequencies are obtained through $\w_{lmn} = \sqrt{k_{lmn}^2 \mu/\rho}$.

\begin{figure}
    \centering
    \includegraphics[width = 0.6\textwidth]{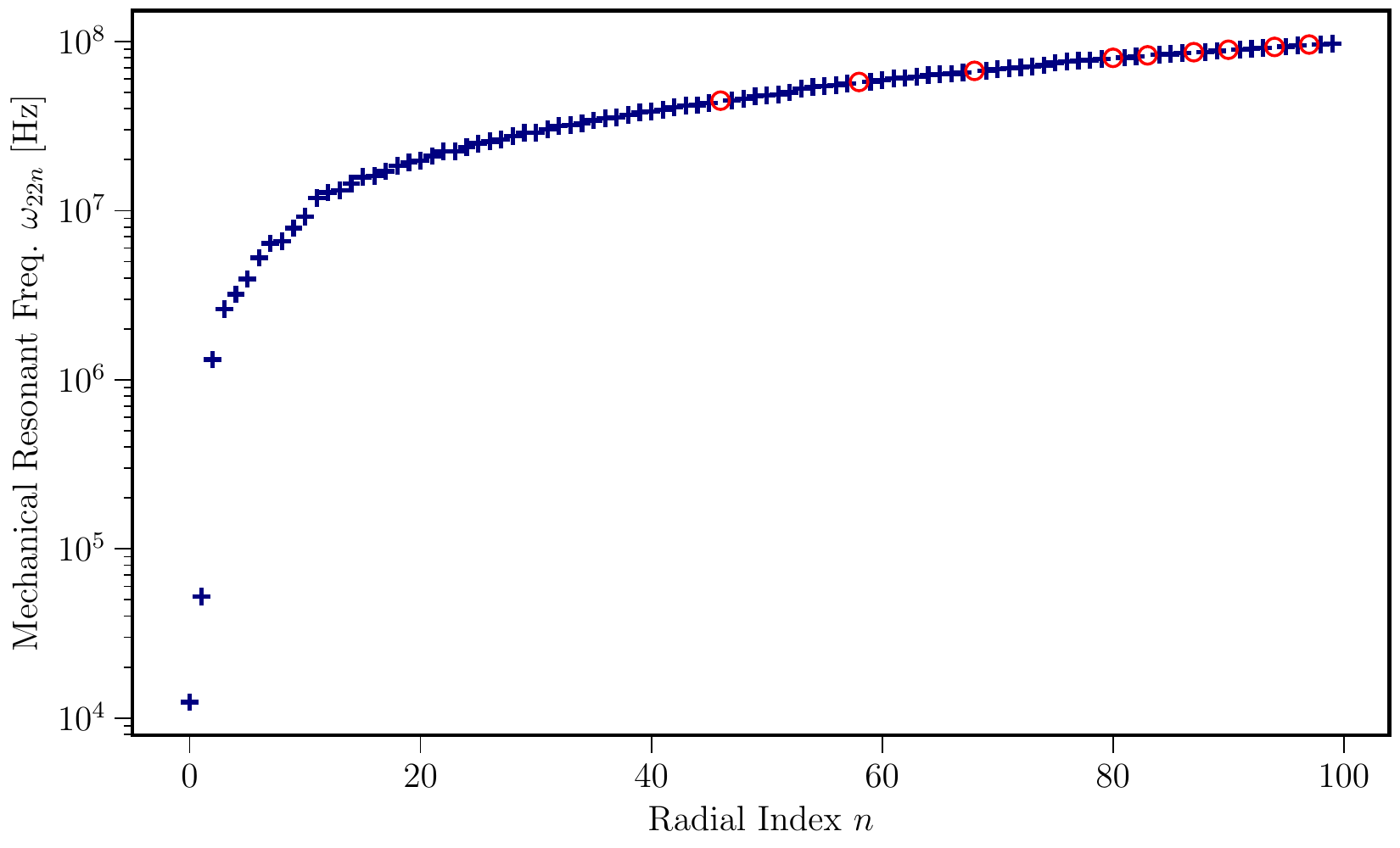}
    \caption{The first 100 mechanical resonant frequencies as a function of the radial index $n$, for $l=m=2$, and with material parameters $\mu = 37.5 \ \text{GPa}$, $\lambda = 147.5 \ \text{GPa}$, and $\rho = 8.57 \ \text{g} / \text{cm}^3$. Red open circles indicate that the corresponding mode is less than one bandwidth $\w_{22n}/Q_{22n}$ away from the $(n+1)^\text{th}$ mode.}
    \label{fig:MechFreqsRadial}
\end{figure}

The resulting frequency spectrum of the spheroidal modes with orbital and azimuthal indices $l=m=2$, for various choices of the radial index $n$, is shown in \Fig{MechFreqsRadial}. The mechanical parameters $\mu = 37.5\ \text{GPa}$, $\lambda = 147.5\ \text{GPa}$, and $\rho = 8.57\ \text{g/cm}^3$ are taken for room temperature and pressure niobium~\cite{10.31399/asm.hb.mhde2.a0003086}. The first few mechanical resonances are sparce in frequency space, but for $n\gtrsim4$, their spacing slowly approaches a continuum where the $n^\text{th}$ and $(n+1)^\text{th}$ frequencies have overlapping bandwidths. We have shown the first $n\leq100$ modes, but it should be noted that for $n\gtrsim 50$, certain modes (marked with red open circles in \Fig{MechFreqsRadial}) are less than one bandwidth away from the next mode. In this case, the response of the material to an external force is no longer well-modeled as the excitation of a single cavity mode. Note that although $m=2$ is held fixed in \Fig{MechFreqsRadial}, the azimuthal index does not affect the frequencies of the resonant modes. Indeed, examining \Eq{eq:MatrixElements}, we see that the zeros of $\det \mathbf{M}$ which fix the wavenumbers $k_{lmn}$  depend on $l$ and are $n$-fold degenerate, but have no $m$-dependence. This is a well-known consequence of spherical symmetry and does not hold for elliptical cavities.

\section{GW-mechanical Overlap Factors}
\label{app:GWmechOverlap}

The full expression for mechanical normal modes of a hollow sphere is given in \Eq{eq:fullMechModeEq}, where spheroidal modes are those with $(c_L^{i,o},\,c_{T_1}^{i,o} \neq 0,~c_{T_2}^{i,o} =0 )$, while toroidal modes are their complement. The overlap factor between GWs and mechanical modes is defined as in \Eq{eq:GWmechcoupling}, which we reproduce here for convenience
\begin{align}
    \eta_{\rm mech}^g = \frac{\hat{h}_{ij}^{\rm TT}}{\Vcav^{1/3} V_{\rm shell}} \int_{V_{\rm shell}} \hspace{-0.3cm} d^3 \mathbf{x} ~ U^{*i}_{p}\, x^j \ .
    \label{eq:GWMechOverlap}
\end{align}
We first show that GWs only couple to mechanical modes with the correct angular momentum structure. The overlap of \Eq{eq:GWMechOverlap} is written in terms of the GW in the TT frame $\hat{h}_{ij}^{\rm TT}$ which is symmetric and traceless. Any such matrix has five independent components, and can be decomposed into a sum over ``helicity" states
\begin{align}
    \hat{h}_{ij}^{\rm TT} = \sum_{m=-2}^2 \hat{h}_m \, \mathcal{Y}_{ij}^{2m} \ ,
\end{align}
where the $\mathcal{Y}_{ij}^{2m}$ are a decomposition of $l=2$ spherical harmonics into an orthogonal basis,
\begin{align}
Y_{2m}(\theta,\p) = \mathcal{Y}_{ij}^{2m} \hat{x}_i \hat{x}_j \ , ~~~
\sum_{i,j} \mathcal{Y}_{ij}^{2m} \left(\mathcal{Y}_{ij}^{2m'}\right)^* = \frac{15}{8\pi} \delta^{mm'} \ ,
\label{eq:sphHarmDecomp}
\end{align}
where $\hat{x}_i$ are unit vectors. This allows us to write the integral in the overlap factor of \Eq{eq:GWMechOverlap} as
\begin{align}
    \sum_{m=-2}^2 \int d^3 \mathbf{x} \, \hat{h}_m\, \mathcal{Y}^{2m}_{ij}\, U^{*i}_{p}\, x^j  = \sum_{m=-2}^2\int d^3 \mathbf{x} \, \hat{h}_m \,Y_{2m}\, \hat{x}_i \,\hat{x}_j \,U^{*i}_{p}\, x^j\ ,
    \label{eq:hqxIntegral}
\end{align}
where we have used the definition of the spherical harmonics of \Eq{eq:sphHarmDecomp}.

Examining first the spheroidal modes, we can write the displacement vector as
\begin{align}
    \mathbf{U}_{lmn} &= c_{0,ln}\left[ f_{ln}(r) Y_{lm}(\theta,\p) \hat{r} - i g_{ln}(r) \hat{r} \times \Lop Y_{lm}(\theta,\p) \right] \ , \label{eq:spheroidalModeSol}
\end{align}
where $f_{ln}(r),~g_{ln}(r)$ encode all radial dependence, including derivatives. From the RHS of \Eq{eq:hqxIntegral}, we can see that \Eq{eq:spheroidalModeSol} will only give a non-zero integral if $l=2$, due to the orthogonality of the spherical harmonics. Turning to the toroidal modes, we can start from \Eq{eq:GWMechOverlap} and write the numerator as 
\begin{align}
   \hat{h}_{ij}^{\rm TT}\int d^3\mathbf{x}\, \,U^{*i}_{p}\, x^j = \hat{h}_{ij}^{\rm TT}\int d^3\mathbf{x}\, \epsilon_{irs} x_j x_r \pd_s \left(F_{ln}(r) Y_{lm}(\theta,\p) \right) \ ,
\end{align}
where $F_{ln}(r)$ encapsulates all radial dependence. The RHS can be integrated by parts, yielding a surface integral term which is zero, since $\epsilon_{irs} x_r \hat{x}_s = 0$, and a volume term,
\begin{align}
     \hat{h}_{ij}^{\rm TT}\int d^3 \mathbf{x}\, \epsilon_{ijr} x_r \left(F_{ln}(r) Y_{lm}(\theta,\p) \right) = 0\ .
\end{align}
The RHS holds since the Levi-Civita symbol is antisymmetric in $i\leftrightarrow j$, while $\hat{h}_{ij}^{\rm TT}$ is symmetric. Thus, GWs do not couple to toroidal modes.

\section{Cavity Perturbation Formalism}
\label{app:Cavity}

In this appendix, we derive how EM modes mix under the influence of quasistatic mechanical deformations of the cavity, following the approach of Refs.~\cite{Meidlinger,Pozar:882338,JONES1964195}. We consider a cavity volume $V'$ that is a small perturbation of the original volume $V$, i.e., $V'=V+\Delta V$. In general, the mode frequencies and profiles can be calculated numerically, but in the limit that the mechanical deformations are small ($\Delta V / V \ll 1$), one can approximately analytically solve for these in terms of the unperturbed quantities. To do this, we begin by approximating the \emph{perturbed} cavity fields $\E$ and $\B$ in terms of the \emph{unperturbed} eigenmodes $\E_n$ and $\B_n$,
\be
\label{eq:fielddecomp}
%\E(\x,t) \simeq \sum_n e_n(t) \, \Big( \E_n(\x) + \Delta \E_n (\xv) \Big)
\E(\x,t) \simeq \sum_n e_n(t) \, \E_n(\x) + \Delta \E (\xv, t)
~~,~~
%\B(\x,t) \simeq \sum_n b_n(t) \, \Big( \B_n(\x) + \Delta \B_n(\xv) \Big) \, ,
\B(\x,t) \simeq \sum_n b_n(t) \, \B_n(\x) + \Delta \B (\xv, t)
~,
\ee
where the dimensionless coefficients $e_n$ and $b_n$ are solely functions of time (their time-dependence will be determined below, which for a perturbed cavity is no longer that of a simple harmonic oscillator).  In the mode expansion above, the sum over $n$ includes solenoidal modes $\grad \cdot \E_n = 0$, while $\Delta \E$ is a small irrotational $\grad \times \Delta \E = 0$ correction to the electric field with corresponding magnetic component $\Delta \B$ (note that only solenoidal cavity modes can be resonantly excited~\cite{condon1941forced,smythe1988static,collin1990field}).
The unperturbed modes satisfy orthogonality conditions when integrated over the unperturbed volume $V$,
\be
\int_V d^3x ~ \E_n \cdot\E_m^* = \int_V d^3x ~ \B_n \cdot\B_m^*=\delta_{nm} \, U_n 
~~,~~
\int_V d^3x ~ \Delta \E \cdot\E_n^* = \int_V d^3x ~ \Delta \B \cdot\B_n^*= 0
~,
\ee
as well as Maxwell's equations $\nabla\times \E_n = -i\w_n \, \B_n$, $\nabla\times \B_n = i\w_n \, \E_n$, where we defined $U_n \equiv \int_{V} d^3 \xv ~ |\E_n|^2 = \int_{V} d^3 \xv ~ |\B_n|^2$ and $\w_n$ are the resonance frequencies of the unperturbed cavity. 

The equations of motion for the coefficients $e_n$ and $b_n$ can be obtained by enforcing the conducting boundary conditions, i.e., $\hat{\n} \times \E \, |_{S^\prime} = 0$ and $\hat{\n} \times \E_n \, |_S = 0$, where $\hat{\n}$ is the unit vector normal to the perturbed $S^\prime$ or unperturbed $S$ cavity surface. The boundary condition for the unperturbed field $\E_n$ can be used to show that
\be
0 = \int_{S} d \A \cdot (\E_n^* \times \B) 
= \int_{V} d^3 \xv ~ \grad \cdot (\E_n^* \times \B)
= \int_{V} d^3 \xv ~ \Big[ (\grad \times \E_n^*) \cdot \B -  \E_n^* \cdot (\grad \times \B) \Big]
~,
\ee
where we used the divergence theorem and a vector calculus identity. Applying Maxwell's equations and decomposing the fields as in \Eq{eq:fielddecomp} then yields 
\be
\label{eq:EMMechEOMen}
%\partial_t e_n \simeq  i \w_n \, b_n + \frac{i}{U_n} \sum\limits_{m} b_m \, \int_{V} d^3 \xv ~ (w_n \, \Delta \B_m \cdot \B_n^*  - \w_m \, \Delta \E_m \cdot \E_n^*) + \order{\Delta V^2}
\partial_t e_n \simeq  i \w_n \, b_n + 
\begin{cases}
\frac{i}{U_n} \sum\limits_{m} b_m \, \int_{\Delta V} d^3 \xv ~ (\w_m \, \E_m \cdot \E_n^*  - \w_n \, \B_m \cdot \B_n^*) + \order{\Delta V^2} & (V^\prime \subset V)
\\
\order{\Delta V^2} & (V \subset V^\prime)
~,
\end{cases}
\ee
where the first and second lines correspond to deformations that either reduce (i.e., concave) or expand (i.e., convex) the cavity volume, respectively (a general volume deformation can be written as a linear combination of these two), and we introduced the appropriate surface currents as dictated by the conducting boundary conditions.\footnote{In the case that $V^\prime \subset V$ or $V \subset V^\prime$, a surface electric current $\K_E \, |_{S^\prime} = \B \times \hat{\n} \, |_{S^\prime}$ on the perturbed surface or a fictitious magnetic current $\K_M \, |_{S} = \E \times \hat{\n} \, |_{S}$ on the unperturbed surface  needs to be introduced in order to satisfy the relevant boundary conditions, respectively.} The boundary condition for the perturbed field $\E$ can be implemented in a similar manner,
\be
0 =  \int_{S^\prime} d \A \cdot (\E \times \B_n^*)
= \int_{V^\prime} d^3 \xv ~ \grad \cdot (\E \times \B_n^*)
= \int_{V^\prime} d^3 \xv ~ \Big[ (\grad \times \E) \cdot \B_n^* - \E \cdot (\grad \times \B_n^*) \Big]
~.
\ee
In addition to using Maxwell's equations and \Eq{eq:fielddecomp}, to simplify the above expression we partition the integral over $V^\prime$ into one over the unperturbed volume $V$ and the deformation $\Delta V$. This then gives  
\begin{align}
\label{eq:EMMechEOMbn}
\partial_t b_n  \simeq  i \w_n  \, e_n  
+ \begin{cases}
\frac{i}{U_n} \,\sum\limits_{m}  \, e_m \, \int_{\Delta V}   d^3 \xv ~ (\w_n \, \E_m \cdot \E_n^* - \w_m \, \B_m \cdot \B_n^* ) + \order{\Delta V^2} & (V^\prime \subset V) 
\\
\frac{2 i}{U_n} \,\sum\limits_{m}  \, e_m \, \int_{\Delta V}   d^3 \xv ~ (\w_n \, \E_m \cdot \E_n^* - \w_m \, \B_m \cdot \B_n^* ) + \order{\Delta V^2} & (V \subset V^\prime)
~.
\end{cases}
%\nl
%&+ \int_{V} d^3 \xv ~ (w_n \, \Delta \E_m \cdot \E_n^*  - \w_m \, \Delta \B_m \cdot \B_n^*) \bigg] 
\end{align}
Taking a time-derivative of \Eq{eq:EMMechEOMen} and using \Eq{eq:EMMechEOMbn} gives
%
%\begin{align}
%\label{eq:EMMechEOM}
%(\partial_t^2 + \w_n^2) \, e_n \simeq \, &\frac{\w_n}{U_n} \,  \sum\limits_{m} \, e_m \, \int_{\Delta V}  \hspace{-0.1cm} d^3 \xv ~ ( \w_m \, \B_m \cdot \B_n^* - \w_n \, \E_m \cdot \E_n^*) 
%\nl
%+ \, &\frac{\w_m^2 - \w_n^2}{U_n} \, \sum\limits_{m} \, e_m \, \int_{V} d^3 \xv ~ \Delta \E_m \cdot \E_n^*  + \order{\Delta V^2}
%~.
%\end{align}
%
\be
\label{eq:EMMechEOM}
(\partial_t^2 + \w_n^2) \, e_n \simeq \, \frac{2}{U_n} \,  \sum\limits_{m} \, e_m \, 
\begin{cases}
\int_{\Delta V} d^3 \xv ~ \Big[ \w_n \, \w_m \, \B_m \cdot \B_n^* - \frac{1}{2} \, (\w_n^2 + \w_m^2) \, \E_m \cdot \E_n^* \Big] 
+ \order{\Delta V^2} & (V^\prime \subset V) 
\\
\int_{\Delta V} d^3 \xv ~ \Big[ \w_n \, \w_m \, \B_m \cdot \B_n^* - \w_n^2 \, \E_m \cdot \E_n^* \Big] 
+ \order{\Delta V^2} & (V \subset V^\prime)
~.
\end{cases}
\ee

\section{Mechanical-EM Overlap Factors}
\label{app:mechanical-em-overlaps}

Mechanical oscillations in the walls of a cavity induce mixing between resonant EM modes of the unperturbed cavity. \Eq{eq:EMMechEOM} shows that the $n^{\text{th}}$ EM mode can be driven by the $m^{\text{th}}$ mode in the perturbed volume $\Delta V$, where $\Delta V$ is determined by the mechanical oscillations $\U_p(\bold{x},t)$ of the cavity walls. Beginning from the results of \App{Cavity}, we will quantify the mixing between the pump mode $\E_0$ and the signal mode $\E_1$ via the mechanical mode $\U_p$.

Including dissipative effects (encapsulated by the signal mode EM quality factor $Q_1$), the equation of motion for the mixing between the pump and signal mode follows directly from \Eq{eq:EMMechEOM},
\be
%\partial_t^2e_1 + \frac{\w_1}{Q_1} \, \partial_t e_1 + \w_1^2 \, e_1 \simeq \frac{\w_1 \, e_0}{\int d^3 \x |\E_1|^2}	\int_{\Delta V}  \hspace{-0.1cm} d^3 \x \, \left(\w_0 \, \B_0 \cdot \B_1^* - \w_1 \, \E_0 \cdot \E_1^*\right) + \frac{(\w_0^2 - \w_1^2) \, e_0}{\int d^3 \x |\E_1|^2} \, \int_{V} d^3 \xv ~ \Delta \E_0 \cdot \E_1^*
\partial_t^2e_1 + \frac{\w_1}{Q_1} \, \partial_t e_1 + \w_1^2 \, e_1 \simeq \frac{2 \, \w_1^2 \, e_0}{\int d^3 \x |\E_1|^2}	\int_{\Delta V}  \hspace{-0.1cm} d^3 \x \, (\B_0 \cdot \B_1^* - \E_0 \cdot \E_1^* ) 
~.
\label{eq:eom_en_dissipation}
\ee
where we took $\w_0 \simeq \w_1$. 
%we can ignore the second term on the RHS above. 
Furthermore,  %the first term on the RHS involving an integral over the perturbed volume can be recast 
recasting the integral over the perturbed volume as a surface integral over the unperturbed surface weighted by the displacement $\U(\x,t) = \sum_p u_p(t) \, \U_p (\x)$, %in which case 
\Eq{eq:eom_en_dissipation} simplifies to  Eqs.~(\ref{eq:EOMem}) and (\ref{eq:etame}).

\subsection{Cavity Setup}

\begin{figure}
    \centering
    \includegraphics[width=0.8\textwidth]{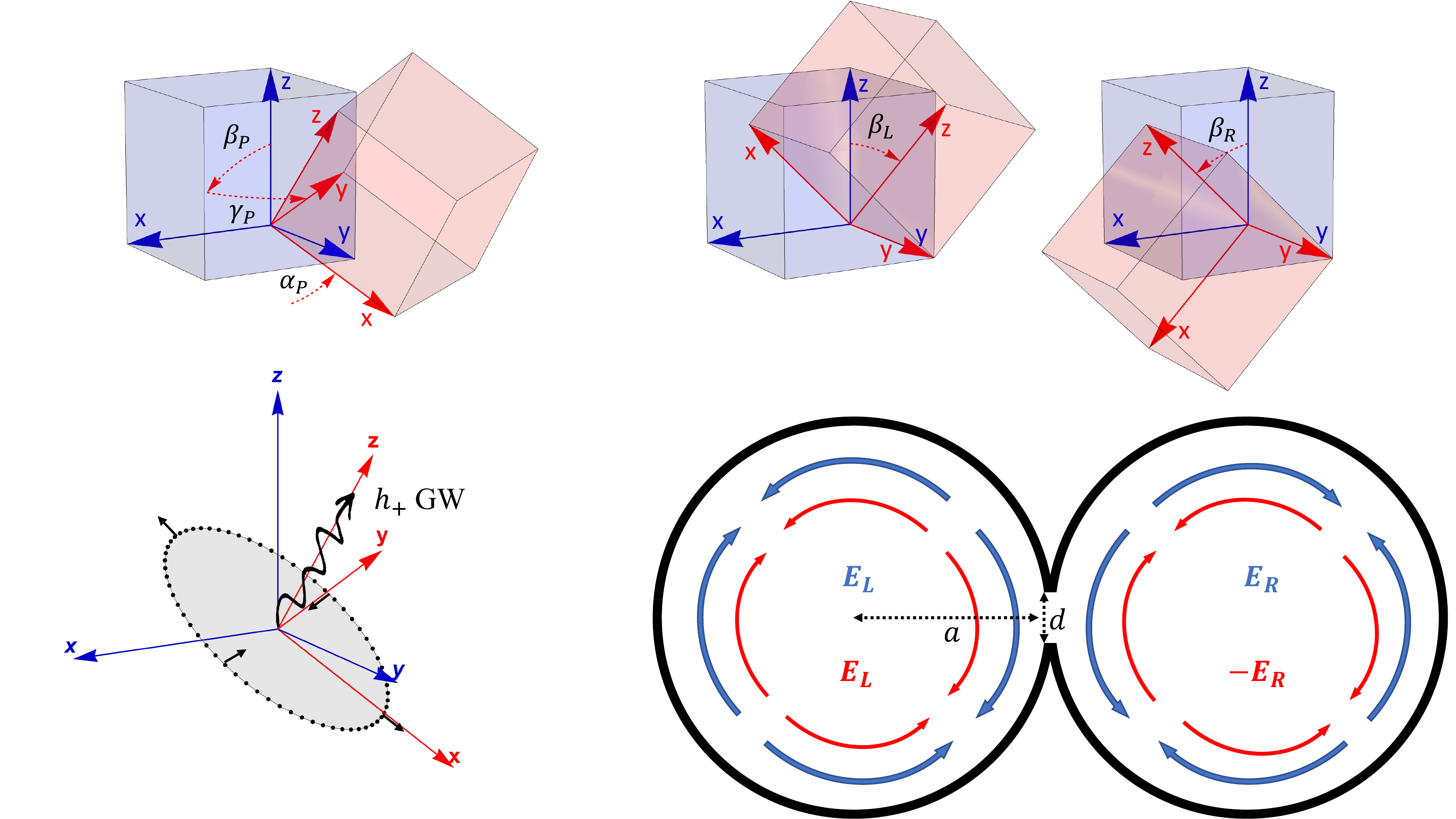}
    \caption{A depiction of the coordinate systems for the mechanical and electromagnetic modes of each cavity-cell and the lab frame. The lab frame is defined relative to astronomical coordinates such that the $x$-axis aligns with $\text{RA}=0^{\circ}$, $\text{dec}=0^{\circ}$, the $y$-axis aligns with $\text{RA}=90^{\circ}$, $\text{dec}=0^{\circ}$, and the $z$-axis aligns with $\text{dec}=90^{\circ}$. \textbf{Top left:} The orientation of the mechanical mode (red axes) is determined by rotating the lab frame (blue axes) with the yaw-pitch-roll matrix $\mathcal{R}_{zyz}(\alpha_p, \beta_p, \gamma_p)$. Since mechanical oscillations are excited by the GW, the yaw-pitch-roll angles can be interpreted in terms of GW parameters. The angle $\alpha_P$ is the angle of the $x$ axis of a plus-polarized GW relative to the blue lab frame, where the plus-polarization is defined by convention to stretch along the $x$ and $y$ mechanical mode axes. The angles $\beta_P$ and $\gamma_P$ are related to the declination and right ascension of the GW source via $\text{RA} = \gamma_p$ and $\text{dec} = \pi/2 - \beta_p$. \textbf{Top right:} Orientation of the electromagnetic mode (red axes), obtained by rotating the lab frame (blue axes) by the yaw-pitch-roll matrices $\mathcal{R}_{zyz}(\alpha_L, \beta_L, \gamma_L)$ and $\mathcal{R}_{zyz}(\alpha_R, \beta_R, \gamma_R)$ for the left and right cells, respectively. For simplicity, only rotations by $\beta_L$ and $\beta_R$ about the laboratory $y$-axis are shown. Physically, the electromagnetic axes are determined by the orientations of the antennas exciting the pump mode. As shown in \Fig{optimal-polarizations} below, the mechanical-electromagnetic coupling $\eta^{\text{EM}}_{\text{mech}}$ is maximized when $|\beta_L - \beta_R| = \pi/2$. \textbf{Bottom left:} A cartoon depiction of a GW corresponding to the mechanical mode orientation shown in the top left. \textbf{Bottom right:} Two-cell pump and signal EM modes with orientations corresponding to the axes shown in the top right.}
 \label{fig:coords}
\end{figure}

A system of two spherical cells each with radius $a$ and coupled via a circular aperture of radius $d \ll a$ is employed, as sketched in \Fig{coords}. Here, we give a detailed description of the resonant modes of the coupled cavity system and discuss the optimal polarizations of the EM modes in the two cells. As discussed in \Sec{modes}, for two nearly identical cells $L$ and $R$, the pump and signal modes, $\E_0$ and $\E_1$, of the coupled system are well-approximated by the symmetric and antisymmetric field configurations of the single cell modes $\tilde{\E}$: $\E_0 = \mathcal{R}_L \tilde{\E} \otimes \mathcal{R}_R \tilde{\E}$ and $\E_1 = \mathcal{R}_L \tilde{\E} \otimes (-\mathcal{R}_R \tilde{\E})$, where $\mathcal{R}_{L,R}$ are rotations acting on the polarization of the left or right cell modes, respectively. The coupling induces a frequency splitting $\Delta \omega_{\text{EM}} \propto (d / a)^3$ between these two modes \cite{PhysRev.66.163}. Since the GW wavelength is assumed to be large relative to the size of the cavity, the two-cell setup is, to a good approximation, not mechanically coupled for frequencies $\wg \gg 1 \ \text{kHz}$.

\begin{figure}
    \centering
    \subfloat{\includegraphics[width=0.45\textwidth]{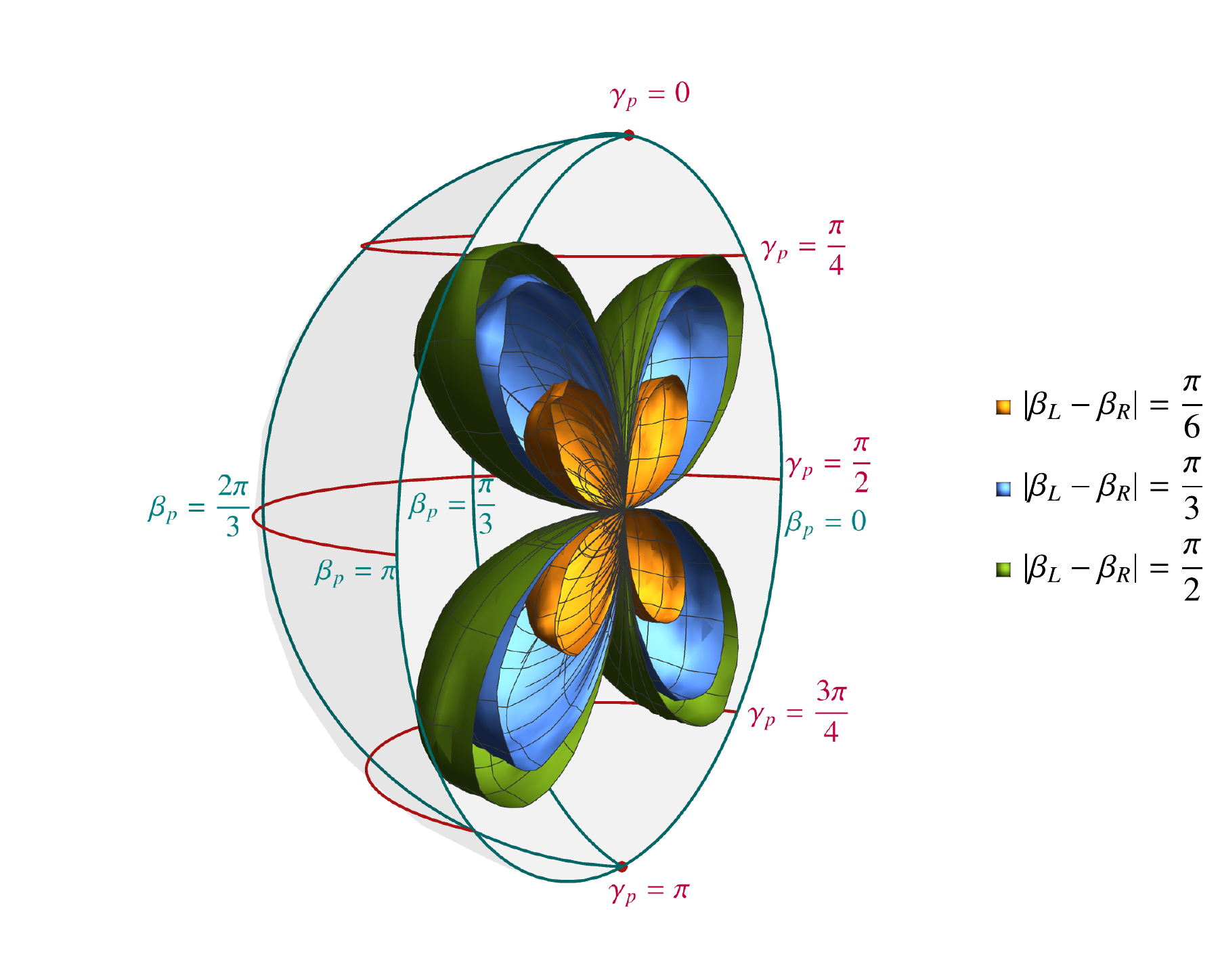}
    }
    ~~\subfloat{\includegraphics[width=0.54\textwidth]{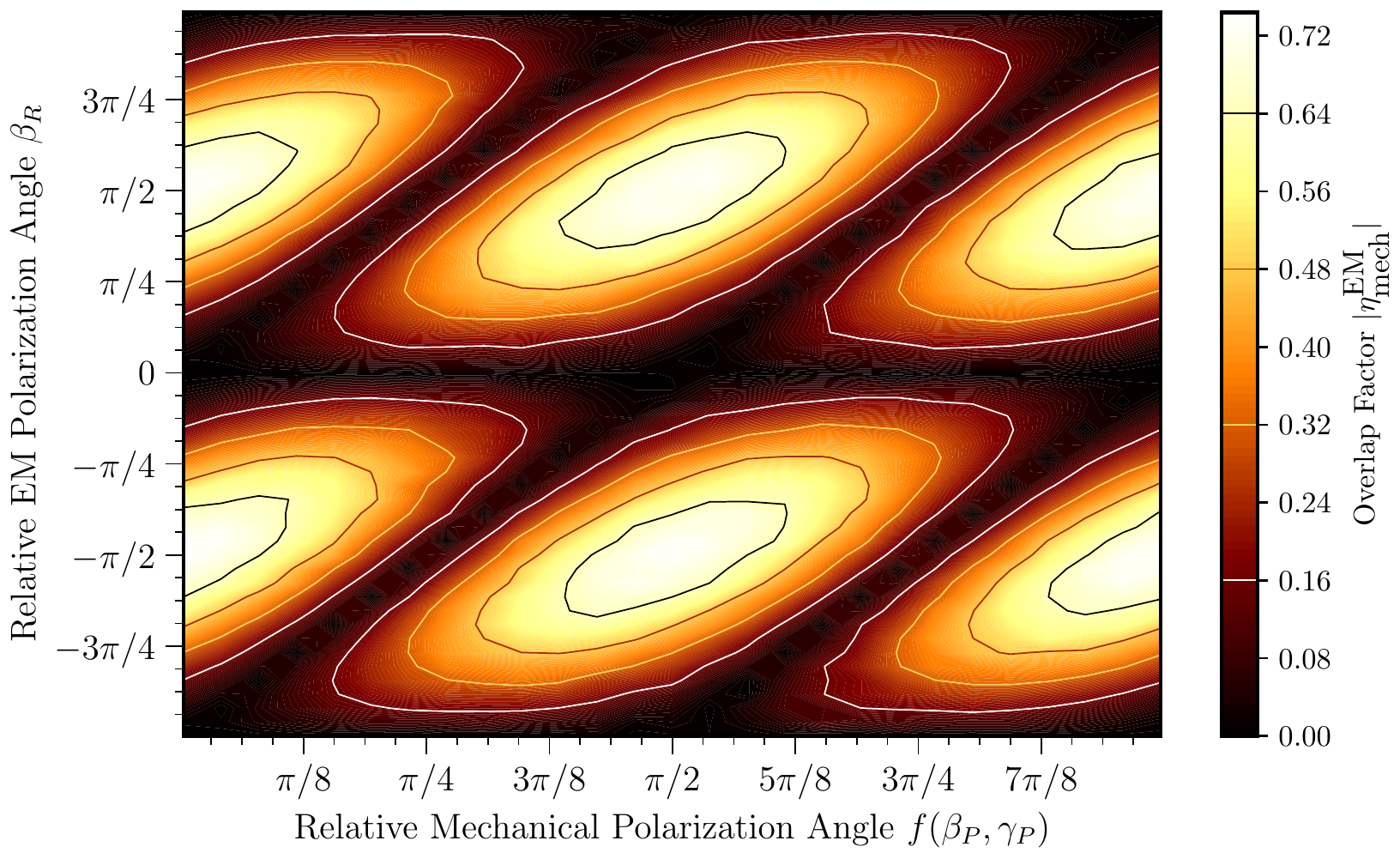}}
    \caption{\textbf{Left:} Spherical plot showing the magnitude of the mechanical-electromagnetic couplings $\eta^{\text{EM}}_{\text{mech}}$ in the radial direction, for a mechanical mode with angular coordinates given by $\beta_p \in [0, \pi]$ and $\gamma_p \in [0, \pi)$. Values are shown for relative electromagnetic polarizations $|\beta_L-\beta_R| \in \{\frac{\pi}{6}, \frac{\pi}{3}, \frac{\pi}{2}\}$. All other rotation angles are set to zero: $\alpha_p = \alpha_L = \gamma_L = \alpha_R = \gamma_R = 0$. \textbf{Right:} Contour plot of $\eta^{\text{EM}}_{\text{mech}}$ for electromagnetic polarization of the right cavity electromagnetic mode relative to the left cavity electromagnetic mode, $\beta_R \in [\frac{-3 \pi}{4}, \frac{3 \pi}{4}]$, and mechanical polarization relative to the left cavity electromagnetic mode, $f(\beta_p, \gamma_p) \in [0, \pi)$. The largest couplings for all GW directions are achieved by setting the relative electromagnetic polarization to $\frac{\pi}{2}$.}
    \label{fig:optimal-polarizations}
\end{figure}

Introducing a second cell breaks the spherical symmetry of the system and gives rise to an additional degree of freedom: the relative polarization of the EM modes in the two cells. Using \Eq{eq:etame}, the coupling between the pump and signal modes can be expressed as a sum of surface integrals over the left and right cells. In general, the expression depends on the relative orientations of the EM modes in the left and right cells and the orientation of the mechanical mode. We specify the orientation of the modes using roll-pitch-yaw matrices $\mathcal{R}_{zyz}(\alpha_i, \beta_i, \gamma_i)$, where the subscript indicates that the first, second, and third arguments rotate about the $z$, $y$, and $z$ axes of the lab frame respectively. The lab frame is taken to be the preferred frame for astronomical coordinates such that the $x$-axis aligns with $\text{RA}=0^{\circ}$, $\text{dec}=0^{\circ}$, the $y$-axis aligns with $\text{RA}=90^{\circ}$, $\text{dec}=0^{\circ}$, and the $z$-axis aligns with $\text{dec}=90^{\circ}$. The physical interpretation of these matrices is described in \Fig{coords}. The mechanical-EM coupling $\eta^{\text{EM}}_{\text{mech}}$ can be written as
\begin{align}
\label{eq:two-cavity-form-factor}
\eta^{\text{EM}}_{\text{mech}} = \frac{V^{1/3}_{\text{cav}}}{\int d^3\x ~ |\tilde{\E}|^2} \Bigg[ &\int_{S_{0 L}}  \hspace{-0.2cm} d \A \cdot (\mathcal{R}_p \U_p)  \,\left( |\mathcal{R}_L \tilde{\E}|^2 - |\mathcal{R}_L \tilde{\B}|^2\right) - \int_{S_{0 R}} \hspace{-0.2cm} d \A \cdot (\mathcal{R}_p \U_p)  \,\left( |\mathcal{R}_R \tilde{\E}|^2 - |\mathcal{R}_R \tilde{\B}|^2\right) \Bigg]
~,
\end{align}
where the relative sign difference between the integrals over the left and right cell surfaces is due to the antisymmetric field configuration of the signal mode. Above, we have also defined
\be
\label{eq:rotated-coords}
\mathcal{R}_p = \mathcal{R}_{zyz}&(\alpha_p,\beta_p,\gamma_p) ~,~
\mathcal{R}_L = \mathcal{R}_{zyz}(\alpha_L,\beta_L,\gamma_L) 
~,~ 
\mathcal{R}_R = \mathcal{R}_{zyz}(\alpha_R,\beta_R,\gamma_R)
\ee
to account for the orientation of the mechanical modes, the EM modes in the left cell, and the EM modes in the right cell relative to the lab coordinate system, respectively. In practice, the EM mode orientation is induced by the way in which the cavity is pumped (e.g., by the placement of an antenna), while the mechanical mode orientation is dictated by the direction of the incoming GW. Importantly, although the coupling $\eta^{\text{EM}}_{\text{mech}}$ is apparently dependent on the polarization of the incoming GW, in practice a superposition of GW polarizations will drive a superposition of mechanical modes with an azimuthal offset of $\pi/4$. In this case, the couplings for the two polarizations (which are identical since \Eq{eq:two-cavity-form-factor} integrates over the entire cavity surface) will simply add in quadrature, yielding the same result as that for a completely polarized GW. Therefore, to be sensitive to as many incoming GW directions as possible, for a given relative EM polarization between the left and right cells, the coupling $\eta^{\text{EM}}_{\text{mech}}$ should be optimized for an $\order{1}$ fraction of all mechanical mode orientations. It is clear from \Eq{eq:two-cavity-form-factor} that if the EM polarizations of the two cells are aligned (i.e., if $\mathcal{R}_L = \mathcal{R}_R)$, then the coupling $\eta^{\text{EM}}_{\text{mech}}$ vanishes. The optimal EM polarization can be determined numerically by computing $\eta^{\text{EM}}_{\text{mech}}$ as a function of the relative EM polarization for all possible mechanical mode polarizations. We find that the optimal EM polarization, as shown in \Fig{optimal-polarizations}, is achieved for a relative orientation of $\pm \pi/2$. In particular, as shown in \Fig{hi-res-couplings-main-body}, for a relative polarization offset fixed to $\pi/2$, $\order{1}$ couplings are achieved for judiciously chosen mechanical and EM modes.

\subsection{Form Factor Scaling with Mechanical Modes}

In \App{mechanical-modes}, we discussed how the frequency of mechanical modes $(\ell=2, n \geq 0, m = 2)$ scales with $n$. Since the noise and signal PSDs at high frequencies are dependent on the mechanical-EM couplings of high $n$ mechanical modes, the scaling of $\eta^{\text{EM}}_{\text{mech}}$ with $n$ is of interest. As shown in \Eq{eq:two-cavity-form-factor}, $\eta^{\text{EM}}_{\text{mech}}$ scales with the radial displacement at the unperturbed surface $\bold{S}$ which, from \Eq{eq:spheroidalModeSol}, is observed to have an angular dependence that is independent of $n$. Thus, form factors differing only by mechanical radial index $n$ will differ only by the ratio of radial displacements at the surface of the unperturbed cavity. Since the radial displacement at the surface does not generically fall off with increasing $n$, we take the EM-mechanical form factor to be independent of $n$.

\section{Mechanical Signal and Noise PSDs}
\label{app:mechPSDs}

We present here the derivation of the mechanical signal and  noise PSDs. The mechanical modes admit a decomposition into spatial- and time-dependent parts, i.e., $\mathbf{U}(\mathbf{x},t) = u_{p}(t) \, \mathbf{U}_{p}(\mathbf{x})$, allowing us to write an equation of motion for the variation of the time-component subject to a force density $\f$
\begin{align}
\pd_t^2 u_{p}(t) + \frac{\w_{p}}{Q_{p}} \pd_t u_{p}(t) + \w_{p}^2 u_{p}(t) = \frac{1}{\Mcav} \int_V d^3\mathbf{x} ~ \mathbf{f}(\mathbf{x},t)\cdot \mathbf{U}_p(\mathbf{x}) \equiv \frac{F_p}{\Mcav} \ .
\label{eq:EoMmechanicalForce}
\end{align}
This equation can be solved in Fourier space to find the PSD of the displacement,
\begin{align}
S_{u_p}(\w) = \frac{S_{F_p}(\w)}{\Mcav^2} \frac{1}{(\w^2-\w_p^2)^2 + (\w \,\w_p/Q_p)^2}
~.
\label{eq:dispPSDgeneric}
\end{align}
As shown in \Eq{eq:EMMechEOM}, such displacements mix EM modes. Solving these equations in Fourier space, we arrive at the PSD for the time-component $e_1$ of the signal EM mode,
\begin{align}
S_{e_1} (\w) = 4 \, |\eta_{\rm mech}^{\rm EM}|^2 \, \Vcav^{-2/3} \, \frac{\w_0^4}{(\w^2-\w_1^2)^2 + (\w\,\w_1/Q_1)^2}  \int\frac{d \w^\prime}{(2\pi)^2} ~ S_{u_p}(\w-\w^\prime) \, S_{e_0}(\w^\prime)
~.
\label{eq:MechSigModePSD}
\end{align}
The PSD of the pump mode $S_{e_0}(\w)$ is determined by the external oscillator used to drive the cavity, which we model as nearly monochromatic, centered near $\w_0$, such that the pump field is $e_0(t) \simeq e_0 \cos{\w_0 t}$. In order to incorporate the width of the external oscillator, $\Delta \w_\text{osc} \lesssim \w_0 / Q_\text{int}$,  the PSD of the pump field is approximate as
\begin{align}
S_{e_0}(\w) \simeq \pi^2 e_0^2 \left(\frac{\Theta(\Delta\w_\text{osc}/2 - |\w-\w_0|)}{\Delta\w_\text{osc}} \right) \ .
\end{align}
Using this in \Eq{eq:MechSigModePSD} as well as accounting for the readout coupling $Q_\text{cpl}$~\cite{Berlin:2020vrk}, we obtain the signal power PSD
\begin{align}
S_{\rm sig}^{\rm (mech)} = 4 \, P_{\rm in} \, \frac{Q_\text{int}}{Q_{\rm cpl}} \, |\eta_{\rm mech}^{\rm EM}|^2 \, \Vcav^{-2/3} \, \frac{\w_0^4}{(\w^2-\w_1^2)^2 + (\w\,\w_1/Q_1)^2} \, \int\frac{d \w^\prime}{4} ~ S_{u_p}(\w-\w^\prime) \, \frac{\Theta(\Delta\w_\text{osc}/2 - |\w'-\w_0|)}{\Delta\w_\text{osc}} \ ,
\label{eq:SigModeVibrations}
\end{align}
where we have defined the input power as $P_{\rm in} = \frac{\w_0}{Q_\text{int}} \, e_0^2 \int d^3 \xv \, |\mathbf{E}_0(\mathbf{x})|^2$. This expression applies for both the displacements induced by the signal or noise. 

As discussed in \Sec{mechsignal}, an incoming GW acts with a force density of $f_i \simeq - R_{i0j0} \, x^j \, \rho_\text{cav}$, which couples solely to spheroidal mechanical modes (see Appendices~\ref{app:mechanical-modes} and \ref{app:GWmechOverlap}) as encapsulated by the coupling-coefficient in \Eq{eq:GWMechOverlap}. The force PSD due to a monochromatic GW is given by
\begin{align}
S^g_{F_{p}} (\w) = \frac{\pi^2}{4}\, \wg^4 \, \Mcav^2 \, \Vcav^{2/3} \, h_0^2 \, |\eta^g_{\rm mech}|^2 \, \big(\delta(\w-\wg) + \delta(\w+\wg)\big) 
~,
\end{align}
which yields a signal power PSD of
\begin{align}
S_\text{sig}^{\text{mech}}(\w) &\simeq \frac{\pi^2}{%16
4} \, P_{\rm in} \, \frac{Q_\text{int}}{Q_{\rm cpl}} \, h_0^2 \, \wg^4 \, |\eta^g_{\rm mech}|^2 \, |\eta_{\rm mech}^{\rm EM}|^2 \, \frac{\w_0^4}{(\w^2-\w_1^2)^2 + (\w\,\w_1/Q_1)^2}
\nl
&\times
\frac{1}{(\wg^2 - \w_{p}^2)^2 + (\wg \w_{p}/Q_{p})^2} \, \frac{\Theta(\Delta\w_\text{osc}/2 - |\w'-\w_0|)}{\Delta\w_\text{osc}}
~.
\label{eq:mechSignalPSD}
\end{align}
In \Sec{comparison}, we presented the transfer function in the case of a spectrally broad GW PSD, $S_h(\w)$. This can be obtained from \Eq{eq:SigModeVibrations} with,
\begin{align}
S^g_{F_{p}}(\w) = \frac{1}{4} \, \wg^4 \, \Mcav^2 \, \Vcav^{2/3} \, |\eta^g_{\rm mech}|^2 \, S_h(\w)
~,
\end{align}
which yields
\begin{align}
S_\text{sig}^{\text{mech}}(\w) \simeq %\frac{1}{16} 
\frac{1}{4} \, P_{\rm in} \, \frac{Q_\text{int}}{Q_{\rm cpl}} \, \frac{\w_0^4 \, |\eta_{\rm mech}^{\rm EM}|^2}{(\w^2-\w_1^2)^2 + (\w\,\w_1/Q_1)^2} \, \frac{\wg^4 \, |\eta^g_{\rm mech}|^2}{((\w-\w_0)^2 - \w_{p}^2)^2 + ((\w-\w_0) \w_{p}/Q_{p})^2} \, S_h(\w-\w_0)
~.
\end{align}
The transfer function $\mathcal{T} = S_\text{sig}^{\text{mech}}(\w_0+\wg) / S_h(\w_g)$ is then easily obtained and is given in \Eq{eq:transferFunction}.

 The calculation for the mechanical noise PSD is obtained from \Eq{eq:MechSigModePSD} in a nearly identical manner. The result is given in \Eq{eq:mechnoisePSD} for the two force PSD benchmarks discussed in \Sec{noise}.

\section{Sources}
\label{app:sources}

In this appendix we discuss possible GW sources in the kHz$-$GHz frequency regime. In particular, we focus on GW sources that are, to a good approximation, monochromatic and have a preferred direction. Since our setup has limited sensitivity to primordial stochastic GW backgrounds, these are not discussed here~\cite{Berlin:2021txa}. 
Instead, we discuss two possible sources of monochromatic GWs: superradiance and primordial black hole inspirals (see also, e.g., Ref.~\cite{Casalderrey-Solana:2022rrn} for an alternative source of high-frequency signals).

\subsection{Black Hole Superradiance}

Gravitational wave signals at high frequencies can arise from black hole superradiance of a new light boson $\p$~\cite{Ternov:1978gq,Zouros:1979iw,Arvanitaki:2009fg,Arvanitaki:2010sy,Detweiler:1980uk,Yoshino:2013ofa,Arvanitaki:2014wva,Brito:2014wla,Brito:2015oca,Zhu:2020tht}. In this scenario, the amplitude of $\p$ is amplified upon  scattering off of a spinning black hole, extracting angular momentum from the black hole in the process.
This bosonic cloud is continuously depleted through annihilation into GWs, resulting in a monochromatic GW signal at a frequency tied to the mass of the host black hole.\footnote{Additional GW signals arise from decays between energy levels of the black hole ``atom," and from bosenovas, but these typically lead to subdominant or short-lived signals~\cite{Arvanitaki:2014wva}.} Superradiance occurs if the boson mass $m_\p$ is less than the angular frequency of the black hole. Ignoring $\order{1}$ dependence on the orbital mode $l$ of the boson cloud, the process requires $\alpha\lesssim \frac{a_*/2}{1+\sqrt{1-a_*^2}}$, where $0\leq a_*\leq 1$ is the dimensionless spin parameter of the black hole and $\alpha \equiv 
G \, M_b \, m_\p \sim 0.2\, ( M_b / M_\odot )\, (m_\p / 10^{-11}\ 
 \text{eV} )$, where $G$ is the gravitational constant and $M_b$ is the black hole mass. For a black hole a distance $D$ away, the expected strain of the produced GWs is~\cite{Aggarwal:2020umq}
\be
h \sim 10^{-23}\,\left(\frac{\Delta a_*}{0.1}\right)\,\left(\frac{1\,{\rm kpc}}{D}\right)\,\left(\frac{M_b}{1\,M_\odot}\right)\,\left(\frac{\alpha}{0.2}\right)^7,
\label{eq:h0_superradiance}
\ee
where $\Delta a_*$ is the parameter quantifying the difference between the initial and final black hole spin. The superradiance condition implies a relation between the mass of the black hole hosting the boson cloud and the mass of the boson, $m_\p \sim 10^{-10}\ \text{eV}  \times (M_\odot/M_b)$, such that \Eq{eq:h0_superradiance} can be recast as a function of $m_\p$. Since the signal arises from annihilation of non-relativistic particles, the frequency of the GWs is $\wg \simeq 2 m_\p$, such that both the expected strain and frequency are directly tied to the mass of the light boson.

We show the possible amplitude of GWs from annihilation of superradiant bosons bound to a black hole $1\,\text{kpc}$ away in \Fig{Reach}. In doing so, we have assumed that $\alpha$ is constant and that $m_\p$ and $M_b$ are therefore dialed accordingly as a function of $m_\p$. We have also normalized $\Delta a_*$ such that the resulting strain matches the careful treatment of the $l=1$ orbital signal shown in Ref.~\cite{Aggarwal:2020umq}. Note that superradiant production of GWs with frequency $\wg \gg 1 \ \text{MHz}$ requires $M_b \ll 1 \ M_\odot$, and thus such black holes must be of primordial origin. Such GWs are typically very coherent. If  self-interactions of $\p$ are negligible, the GW frequency drift rate is $\dot{\w}_g = 7 \times 10^{-15}\ \text{Hz}\ \text{sec}^{-1} \times (\alpha / 0.1)^{17}\, (\w_g / \text{kHz})^2$~\cite{Baryakhtar:2020gao}. The effective quality factor of the signal is the number of oscillations it takes for this drift to result in an $\order{1}$ phase, i.e., $Q_g \sim \wg / \sqrt{\dot{\w}_g} \sim 10^{10} \times (0.1 / \alpha)^{17/2}$. Equivalently, for $\alpha = 0.1$, $\w_0 \sim 1 \ \text{GHz}$ and $Q_1 \sim 10^5$, the GW signal remains in the cavity bandwidth for a time of $t \sim 10^5 \ \text{yrs} \times (\text{MHz} / \wg)^2$, such that the signal is coherent over timescales much longer than the typical  experimental integration time.

\subsection{Mergers of Compact Objects}

High-frequency GWs may also arise from the inspiral of two sub-solar mass compact objects~\cite{Maggiore}. The GW frequency increases during the inspiral and is largest close to the merger of the two objects. The signal coherence time $\tau_b$ can be defined as the duration the GW frequency stays within the cavity bandwidth, 
\begin{eqnarray}
	\tau_b \sim 10^{-3} \,{\rm s}\,\left(\frac{10^5}{Q}\right)\left(\frac{10^{-11}\,M_\odot}{M_b}\right)^{5/ 3}\left(\frac{1\,{\rm GHz}}{\omega_g}\right)^{8 / 3}
~,
\end{eqnarray}
where we assumed that both inspiraling objects have the same mass $M_b$. To fully ring up the cavity we require that the time it takes to sweep over the cavity bandwidth is longer than the ring up time, which is $t_\text{ring} \sim \text{GHz} / Q_1$ for GW frequencies much greater than the lowest-lying mechanical resonance ($\wg \gg 10 \ \text{kHz}$). This leads to the requirement
\begin{eqnarray}
	M_b <
 5\times 10^{-11}\ M_\odot \ \left( \frac{10^5}{Q_1} \right)^{6/5}  \left( \frac{1 \ \text{GHz}}{\wg} \right)^{8/5}
~.
	\label{eq:ringup_condition}
\end{eqnarray}
Thus, compact objects with $M_b\lesssim 10^{-10} \ M_\odot$ generically fully ring up the cavity for $\wg \sim 10 \ \text{kHz}- 1 \ \text{GHz}$. For larger masses of $M_b \sim 10^{-8} \ M_\odot$ or $M_b \sim 10^{-4} \ M_\odot$, the cavity is only fully rung up for frequencies $\w_g \lesssim 30 \ \text{MHz}$ and  $\w_g \lesssim 100 \ \text{kHz}$, respectively. The expected strain from inspirals a distance $D$ away is roughly~\cite{Maggiore}
\be
h_0 \sim 10^{-29} \times \bigg(\frac{1\ \text{pc}}{D}\bigg)\bigg( \frac{M_b}{10^{-11}\, M_\odot}\bigg)^{5/3} \bigg( \frac{\omega_g}{1 \ \text{GHz}} \bigg)^{2/3}
~.
\label{eq:strain_PBH_inspiral}
\ee

\section{Optimization of Sensitivity}
\label{app:Optimal}

\begin{figure}
    \centering
    \includegraphics[width = 0.6\textwidth]{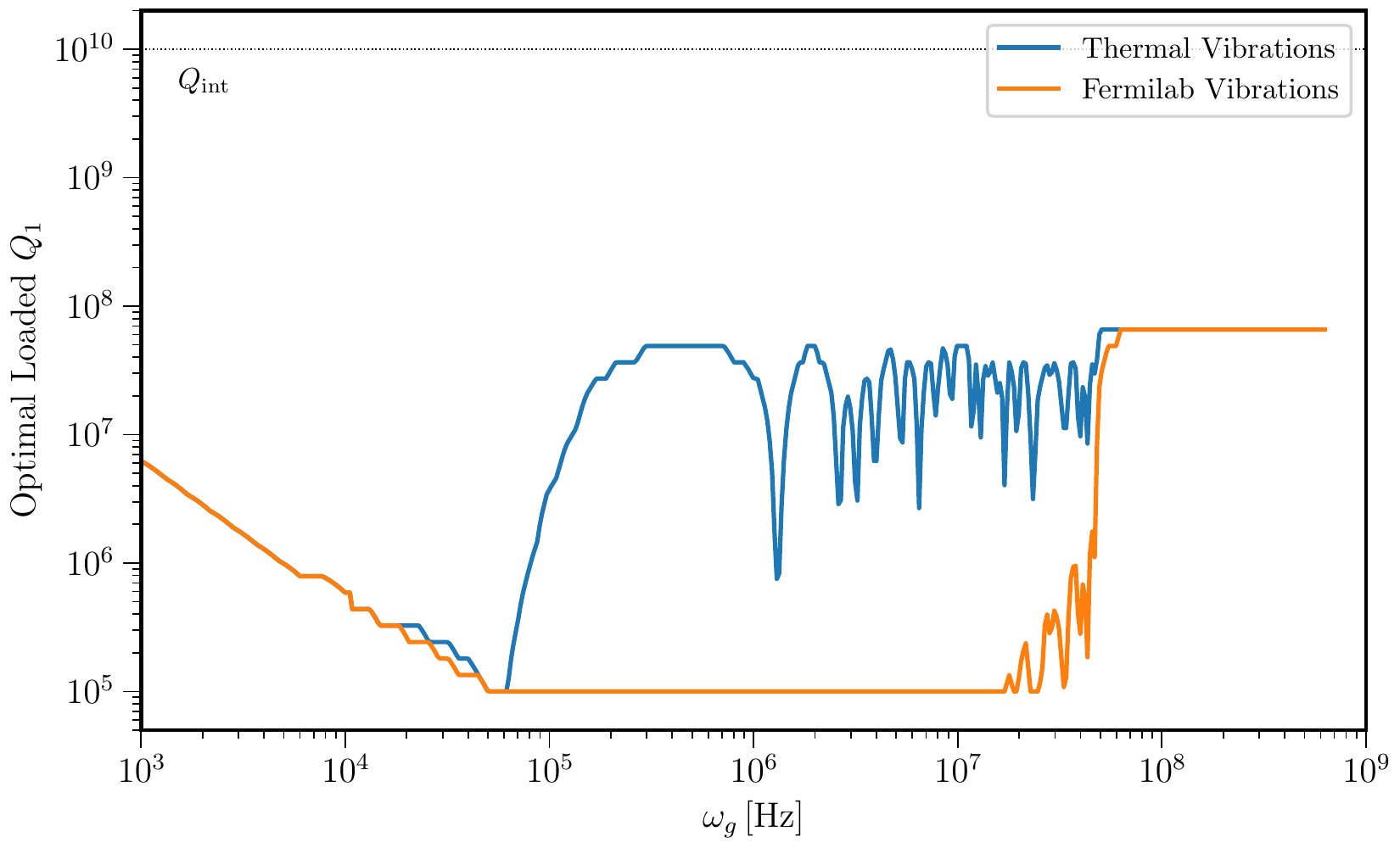}
    \caption{The optimal loaded quality factor of the signal mode, $Q_1(\wg)$, calculated numerically for a scanning strategy. The blue line corresponds to the optimal $Q_1$ when mechanical mixing noise comes from thermal vibrations, while the orange assumes a flat force PSD giving a $1\ \Hz$ fluctuation of the EM resonant frequencies~\cite{fnalex}. Features are explained in \App{Optimal}. The intrinsic cavity quality factor is $Q_{\rm int} = 10^{10}$, and is shown with a dashed line. For the broadband strategy, the optimal $Q_1 = 10^5$.}
    \label{fig:OptimalQLoaded}
\end{figure}

As discussed in \Sec{sensitivity}, the sensitivity to GWs can be improved by overcoupling to the signal mode, lowering the loaded quality factor $Q_1 \leq Q_\text{int}$. The result of a numerical optimization requiring that $10^5 \leq Q_1 \leq 10^{10}$ is shown in \Fig{OptimalQLoaded}, assuming the scanning strategy discussed in \Sec{sensitivity}. The integration time at each frequency-step is given by $t_{\rm int} \sim t_e \, \text{min} \big( \w_1 / (Q_1 \, \wg) \, , \, 1 \big)$ for a fixed $e$-fold time of $t_e = 1\, \text{yr}$. The blue line in \Fig{OptimalQLoaded} shows the optimal $Q_1$ assuming that mechanical noise is attenuated to its irreducible thermal value, corresponding to the ``scanning (thermal)" curve in Figs.~\ref{fig:Reach} and \ref{fig:SHReach}. In orange, we have shown the optimal $Q_1$ assuming instead vibrational noise as inferred by recent measurements of cavity microphonics at Fermilab~\cite{DarkSRF,Pischalnikov:2019iyu}, corresponding to the ``scanning" curves in Figs.~\ref{fig:Reach} and \ref{fig:SHReach}. At high frequencies, $\wg \gtrsim 10^8\ \text{Hz}$, EM thermal noise dominates for both cases, and the optimal loaded quality factor is approximately $Q_1\sim Q_{\rm int}\, (\w_1/T) \sim 10^8$~\cite{Chaudhuri:2018rqn,Berlin:2019ahk}. At low frequencies $\wg \lesssim 10^5\ \text{Hz}$, the optimization of $Q_1$ requires balancing mechanical and amplifier noise, leading to the scaling of $Q_1 \propto 1/\wg$ as shown in \Fig{OptimalQLoaded}. More generally, the preferred value of $Q_1$ depends on the dominant noise source. In a mechanical-noise-limited setup, overcoupling as much as possible is beneficial as long as doing so does not lead to another noise source dominating. This is seen most clearly along the orange curve in \Fig{OptimalQLoaded}, where mechanical noise is dominant for GW frequencies of $10^5\ \text{Hz} \lesssim \wg \lesssim 10^8\ \text{Hz}$ and $Q_1 \sim 10^5$ is preferred. Instead, along the blue curve, mechanical noise dominates over EM thermal noise in the same frequency range only near  individual mechanical resonances (see the left panel of \Fig{PSDs}). As a result, we observe that near mechanical resonances $Q_1 \sim 10^5$ and away from mechanical resonances $Q_1 \sim Q_{\rm int}\, \w_1/T \sim 10^8$. The exact optimal value of $Q_1$ near mechanical resonances is not always shown due to the coarse-graining of the numerical optimization procedure.

For a broadband setup corresponding to the ``non-scanning (thermal)" curves in Figs.~\ref{fig:Reach} and \ref{fig:SHReach}, the reach is optimized for $Q_1 = 10^5$. Since no scanning is involved, the experimental integration time is independent of $Q_1$, allowing us to understand the $Q_1$-scaling of the GW sensitivity entirely from the ratio of signal and noise PSDs. We can therefore examine the noise-equivalent strain $S_h^{\rm noise}$, which is a proxy for this ratio, to understand the scaling. In a broadband setup employing a fixed EM frequency splitting, mechanical noise dominates for GW frequencies $\w_g \lesssim \w_1 - \w_0 \sim 10 \ \text{kHz}$. From \Eq{eq:shmechnoise}, $S_h^{\rm noise}$ and therefore the sensitivity, is independent of $Q_1$. Instead, for $\wg \gtrsim \w_1 - \w_0 \sim 10 \ \text{kHz}$, amplifier noise dominates. Since amplifier noise is external to the cavity, it is independent of $Q_1$, such that the $Q_1$-dependence of $S_h^{\rm noise}$ is entirely due to the transfer function $\mathcal{T}$ (see \Eq{eq:shAmpNoise}). As a result, it is optimal to overcouple as much as possible, i.e, $Q_1 \simeq 10^5$.

\bibliography{bibliography}
\end{document}